\documentclass[manuscript]{aastex}

\shorttitle{Haumea family}
\shortauthors{Thirouin et al.}

\begin{document}


\title{Rotational properties of the Haumea family members and candidates: Short-term variability.}


\author{Audrey Thirouin\altaffilmark{1}}
\affil{Lowell Observatory, 1400 W Mars Hill Rd, Flagstaff, Arizona, 86001, United States of America. }
\email{thirouin@lowell.edu}

\author{Scott S. Sheppard\altaffilmark{2}}
\affil{Department of Terrestrial Magnetism (DTM), Carnegie Institution for Science, 5241 Broad Branch Rd. NW, Washington, District of Columbia, 20015, United States of America.}

\author{Keith S. Noll\altaffilmark{3}}
\affil{NASA Goddard Space Flight Center (NASA-GSFC), Greenbelt, Maryland, 20771, United States of America.}

\author{Nicholas A. Moskovitz\altaffilmark{1}}
\affil{Lowell Observatory, 1400 W Mars Hill Rd, Flagstaff, Arizona, 86001, United States of America. }

\author{Jos\'e-Luis Ortiz\altaffilmark{4}}
\affil{Instituto de Astrof\'{\i}sica  de Andaluc\'{\i}a (IAA-CSIC), Apt 3004, 18080,  Granada, Spain.}

\and

\author{Alain Doressoundiram\altaffilmark{5}}
\affil{Observatoire de Paris-LESIA, 5 Place Jules Janssen, Meudon Cedex 92195, France.}




\begin{abstract}
Haumea is one of the most interesting and intriguing transneptunian objects (TNOs). It is a large, bright, fast rotator, and its spectrum indicates nearly pure water ice on the surface. It has at least two satellites and a dynamically related family of more than ten TNOs with very similar proper orbital parameters and similar surface properties. The Haumean family is the only one currently known in the transneptunian belt. Various models have been proposed but the formation of the family remains poorly understood. In this work, we have investigated the rotational properties of the family members and unconfirmed family candidates with short-term variability studies, and report the most complete review to date. We present results based on five years of observations and report the short-term variability of five family members, and seven candidates. The mean rotational periods, from Maxwellian fits to the frequency distributions, are 6.27$\pm$1.19~h for the confirmed family members, 6.44$\pm$1.16~h for the candidates, and 7.65$\pm$0.54~h for other TNOs (without relation to the family). According to our study, there is a suggestion that Haumea family members rotate faster than other TNOs, however, the sample of family member is still too limited for a secure conclusion. We also highlight the fast rotation of 2002~GH$_{32}$. This object has a 0.36$\pm$0.02~mag amplitude lightcurve and a rotational period of about 3.98~h. Assuming 2002~GH$_{32}$ is a triaxial object in hydrostatic equilibrium, we derive a lower limit to the density of 2.56~g cm$^{-3}$. This density is similar to Haumea's and much more dense than other small TNO densities. 
\end{abstract}





\section{Introduction}

The dwarf planet Haumea, (136108) 2003~EL$_{61}$, has been well observed since its discovery and exhibits a number of interesting and unusual characteristics: 
\begin{itemize}
\item bright with a visual magnitude about 17.
\item large with a diameter around 1200~km and a geometric albedo of 0.70-0.75 \citep{Lellouch2010}.
\item a very elongated shape and a dark spot \citep{Lacerda2008}.
\item fast rotational period around 3.9~h \citep{Rabinowitz2006, Lacerda2008, Thirouin2010}. 
\item high density of 2.5-3.3~g cm$^{-3}$ \citep{Rabinowitz2006, Thirouin2010}.
\item pure water ice surface \citep{Trujillo2007, Merlin2007, Tegler2007}.
\item at least two satellites: Namaka and Hi'iaka \citep{MPC_FirstSatellite_Haumea,MPC_SecondSatellite_Haumea,Brown2006}.
\item ten objects (plus Haumea and its satellites) have similar proper orbital parameters and similar surface properties. 
\end{itemize} 

\citet{Noll2005} noted that three objects, (19308) 1996~TO$_{66}$, (24836) 1995~SM$_{55}$, and (86047) 1999~OY$_{3}$ had unusually blue colors in a \textit{Hubble Space Telescope} (HST) survey of 81 TNOs using NICMOS (F110W-F160W color). The authors suggested that these unusual colors could be due to bodies covered by relatively fresh ices. 
\citet{Brown2007} further suggested that these three objects with several others had similar proper orbital parameters and surface properties to Haumea. This lead to the idea that Haumea, its satellites, and these other objects formed a family. \citet{Brown2007} proposed that the proto-Haumea\footnote{The term "proto-Haumea" is used to refer to the object prior to the formation of the family. The name "Haumea" is used to refer to the actual object.} suffered a catastrophic impact that ejected a large fraction of its icy mantle, which formed the two satellites and the dynamical family. \citet{Levison2008} found that the Haumean family is likely the only collisional family in the Trans-Neptunian belt. However, \citet{Marcus2011} and \citet{CampoBagatin2012} have argued that there could be more families in this region. They estimated that a collision on a 400~km body would have produced a largest fragment not smaller than $\sim$300~km and fragments in the 50-100~km size range (23-24~mag) making their identification difficult with existing surveys (e.g. \citet{Brown2015, Sheppard2011}).  

In this work, we report observations of short-term variability of Haumea family members and candidates, including objects not previously observed. We performed CCD photometric observations using several telescopes in Spain and the United States over a period of five years. We report the short-term variability of twelve objects. We compare the rotational properties of the Haumea family members to non-family transneptunian objects (TNOs). As a test of whether the peculiar creation and history of this group of object, may have resulted in rotational properties that are different from those of other TNOs. We examine also the lightcurve amplitude distribution of the family members and candidates. Body elongation and lower limit to the density are derived from lightcurves.     

This paper is divided into six sections. In the next section, we review the Haumea family and define terminology. Then, we describe the observations and the data set analyzed. In Section 4, we present our main results for each object. In Section 5, we discuss our results and present a summary about the Haumea family members and candidates. Finally, Section 6 is dedicated to the conclusions of this work.   


\section{Haumea family members}


The Haumea family is composed of objects sharing similar proper orbital elements and similar surface properties \citep{Brown2007}. Currently, the list of confirmed Haumea family members is: 
\begin{itemize}
\item \citet{Brown2007} identified (24835) 1995~SM$_{55}$, (19308) 1996~TO$_{66}$, (55636) 2002~TX$_{300}$, (120178) 2003~OP$_{32}$, (145453) 2005~RR$_{43}$, (136108) Haumea, Namaka, and Hi'iaka. 
\item \citet{Ragozzine2007} added (86047) 1999~OY$_{3}$, and 2003~UZ$_{117}$. 
\item \citet{Schaller2008} added (308193) 2005~CB$_{79}$. 
\item \citet{Snodgrass2010} included 2003~SQ$_{317}$\footnote{\citet{Snodgrass2010} considered this object only as a \textit{probable} member due to the lack of optical colors and relatively large uncertainty on their (J-H$_{S}$) value (see \citet{Snodgrass2010} for more details). Therefore, the water ice detection is only preliminary. According to \citet{Lacerda2014}, this object has a nearly solar surface colour matching colours of the other members, but its phase function is much steeper. In order to provide the most complete study, we will consider this object as member. }. 
\item \citet{Trujillo2011} confirmed the membership of 2009~YE$_{7}$.
\end{itemize} 

\citet{Ragozzine2007} published a list of possible family members (hereinafter, candidates) that have proper orbital elements consistent with the family but without any near infrared spectra confirming the presence of water ice on their surfaces (see \citet{Ragozzine2007} for a more complete definition of the candidate sample). Some of these candidates have been rejected by \citet{Snodgrass2010}, and \citet{Carry2012} based on colour studies. 
Recently, Sheppard and Trujillo discovered another candidate in their survey dedicated to the search of Sedna-like objects \citep{Trujillo2014, Sheppard2015}. This object, 2014~FT$_{71}$, has proper orbital elements consistent with Haumea's, but color/spectroscopic observations needed to confirm family membership are lacking \citep{Sheppard2015FT71}.    

\citet{Volk2012} suggested that (315530) 2008~AP$_{129}$ belongs to a new class of rockier family members. This object has proper elements consistent with being a member of the family but does not have a strong water ice signature \citep{Brown2012}. \citet{Volk2012} speculated that this object could be a fragment from an inner part of a differentiated proto-Haumea. On the other hand, \citet{Cook2011} based on \citet{Desch2009} work suggested that the proto-Haumea was only partially differentiated. In fact, \citet{Desch2009} showed that TNOs with radii in the range 500-1000~km are only partially differentiated with a rocky core and an icy mantle surrounded by a thick crust of rock/ice mixture. Such a crust never reached temperatures high enough to melt or differentiate. In that case, the fragments forming the Haumea family are from the icy mantle and the crust, and so, one might expect a mix of icy and rocky members in the family. As 2008~AP$_{129}$ appears as a "transition object" between the icy and rocky population, for the purpose of this work, we will consider it as a candidate. Orcus and Pluto-Charon have also water ice on their surface, but, because their proper orbital elements are not similar to the rest of the family and candidates, they will not be considered in this work \citep{Fornasier2004, Trujillo2005, Carry2011, Cruikshank2015}.  
 




\section{Observations and data reduction}
\subsection{Runs and Telescopes}

We present data obtained with Lowell Observatory's 4.3~m Discovery Channel Telescope (DCT), the 3.58~m Telescopio Nazionale Galileo (TNG), the 2.5~m Isaac Newton Telescope (INT), the 2.2~m Centro Astron\'{o}mico Hispano Alem\'{a}n (CAHA) telescope at Calar Alto Observatory, and the 1.5~m Sierra Nevada Observatory (OSN) telescope between 2011 and 2015.

The DCT is forty miles southeast of Flagstaff at the Happy Jack site (Arizona, United States of America). Images were obtained using the Large Monolithic Imager (LMI) which is a 6144$\times$6160 CCD. The total field of view is 12.5$\arcmin$$\times$12.5$\arcmin$ with a pixel scale of 0.12$\arcsec$/pixel (unbinned). Images were obtained using the 3$\times$3 binning mode.

The TNG and INT are located at the Roque de los Muchachos Observatory (La Palma, Canary Islands, Spain). Images were obtained using the Device Optimized for the LOw RESolution instrument (DOLORES or LRS). This device has a camera and a spectrograph installed at the Nasmyth B telescope focus. We observed in imaging mode and a 2$\times$2 binning mode. The camera is equipped with a 2048x2048 CCD with a pixel size of 13.5$\micron$. The field of view is 8.6$\arcmin$$\times$8.6$\arcmin$ with a 0.252$\arcsec$/pixel scale (pixel scale for a 1$\times$1 binning). Our observations with the INT were obtained with the Wide Field Camera (WFC) instrument. This camera consists of 4 thinned EEV 2154$\times$4200 CCDs for a total field of view of 34$\arcmin$$\times$34$\arcmin$. The pixel scale is 0.33$\arcsec$/pixel (pixel scale for a 1$\times$1 binning). 
 
The 2.2~m CAHA telescope at Calar Alto Observatory is located in the Sierra de Los Filabres (Almeria, Spain). 
We used the Calar Alto Faint Object Spectrograph (CAFOS) instrument located at the Cassegrain focus of the telescope. CAFOS is equipped with a 2048$\times$2048 pixels CCD and the image scale is 0.53$\arcsec$/pixel (pixel scale for a 1$\times$1 binning). Images were obtained using 2$\times$2 binning. 

The 1.5~m telescope is located at the Observatory of Sierra Nevada (OSN), at Loma de Dilar in the National Park of Sierra Nevada (Granada, Spain). Observations were carried out by means of a 2k$\times$2k CCD, with a total field of view of 7.8$\arcmin$$\times$7.8$\arcmin$. We used a 2$\times$2 binning mode, which changes the image scale to 0.46$\arcsec$/pixel.

\subsection{Observing strategy, Data reduction and analysis}
 
Exposure times were chosen based on two main factors: i) exposure time had to be long enough to achieve a signal-to-noise ratio (S/N) sufficient to study the observed object (typically, S/N$>$20); ii) exposure time had to be short enough to avoid elongated images. We always tracked the telescope at sidereal speed. The drift rates of TNOs are low, typically $\sim$2$\arcsec$/h, so exposure times around 200 to 900 seconds were used.

Observations at the OSN were performed without filter in order to maximize the S/N. As the main goal of our study is short-term variability via relative photometry, the use of unfiltered images without absolute calibration is not a problem. The r' Sloan filter was used during our observations with the TNG. With the 2.2~m CAHA telescope, we used the Schott KG1 filter (near-infrared blocking filter), and the VR filter (broad-band filter) at the DCT. Such filters were chosen to maximize the object S/N and to minimize the fringing that appears at longer wavelengths in these instruments. 

In this work, we focused on five confirmed members of the family: 1995~SM$_{55}$, 1999~OY$_{3}$, 2003~OP$_{32}$, 2003~UZ$_{117}$, and 2009~YE$_{7}$, and seven candidates: 1999~CD$_{158}$, 2000~CG$_{105}$, 2002~GH$_{32}$, 2003~HA$_{57}$, 2003~HX$_{56}$, 2005~GE$_{187}$, and 2008~AP$_{129}$. All relevant geometric information about the observed objects at the date of observation, the number of images and filters are summarized in Table~\ref{Log_Obs}. 








We used the standard data calibration and reduction techniques described in \citet{Thirouin2010, Thirouin2012, Thirouin2014}. 

\subsection{Period-detection methods}
\label{sec:singledouble}
 
The time-series photometry of each target was inspected for periodicities by means of the Lomb technique \citep{Lomb1976} as implemented in \cite{Press1992}. We also checked our results with several other time-series analysis techniques, such as Phase Dispersion Minimization (PDM) \citep{Stellingwerf1978}, and CLEAN technique \citep{Foster1995}. \cite{Harris1989} method and its improvement \citep{Pravec1996} was also used (hereinafter Pravec-Harris method). Finally, in order to measure the full amplitude (or peak-to-peak amplitude) of short-term variability, a first or second order Fourier fit (depending if we are considering a single- or double-peaked rotational periodicity) to the data has been performed.    

We must point out that to distinguish between shape and/or albedo contribution(s) in a lightcurve is not trivial. In fact, care has to be taken to distinguish between a single- or double-peaked lightcurve (see Fig~1 and discussion related in \citet{Thirouin2014}). Except for a pole-on view of an object, in which no rotational variability can be observed, the observer will detect rotational variability for the rest of configurations of the spin axis. Assuming a triaxial ellipsoid (Jacobi ellipsoid), we have to expect a lightcurve with two maxima and two minima, corresponding to a full
rotation (a double-peaked lightcurve). However, if the object is spherical or oblate (MacLaurin spheroid) without any albedo variation on its surface, we have to expect a flat lightcurve. If this spheroid presents albedo variation on its surface, we have to expect a lightcurve with one maximum and one minimum (i.e., a single-peaked lightcurve). \\

When the lightcurve amplitude is small, it is very difficult or even impossible to distinguish if the lightcurve is single- or double-peaked. Therefore, we have to find a criterion to distinguish between both cases. In \citet{Thirouin2010, Duffard2009}, we proposed a threshold at 0.15~mag to distinguish between shape and albedo effects. This criterion is a simplification because there may be elongated objects whose rotational variability is smaller than 0.15~mag simply because their rotation axes are viewed close to pole-on from Earth. However, these objects are only a small fraction because statistically only very few objects have spin axes near the pole-on orientation\footnote{In case of pole-on observation, even if the object is very elongated, lightcurve will be flat. However, the probability of such a case is low. In fact, Equation 6a of \citet{Lacerda2003} estimates the probability to observe an object with a pole-on orientation. Assuming an angle $\theta$=5$^\circ$, the probability to see an object with a pole-on orientation$\pm$5$^\circ$ is less than 1~$\%$.}. In \citet{Thirouin2013}, we tested what is the lightcurve amplitude limit to distinguish between shape- and albedo-dominated lighcurves (i.e. to distinguish between single- and double-peaked lighcurves). We tested three lightcurve amplitude limits: i) a threshold at 0.10~mag, ii) at 0.15~mag, and iii) at 0.20~mag, to distinguish between single- and double-peaked lightcurves. The best fit (i.e. with the highest confidence level) has been obtained considering a lightcurve amplitude limit of 0.15~mag. Such a criterion with a threshold at 0.15~mag has been used already by several investigators as the transition from low to medium variability \citep{Lacerda2006, Sheppard2008}. In the asteroid case, albedo variations are usually responsible for lightcurve amplitudes between 0.1~mag and 0.2~mag \citep{Magnusson1991, Lupishko1983, Degewij1979}. \\
Albedo variation in the asteroid modeling is negligible but not in the TNO case. The shape dominates in the asteroid case, but not for the TNOs \citep{Lacerda2008}. \citet{Lacerda2008} simulated a synthetic ellipsoidal object without any mark of albedo on its surface, and the result is a perfectly symmetric double-peaked lightcurve. But if a spot or a hemispheric albedo variation is present on the surface of the same ellipsoidal object, the result is an asymmetric double-peaked lightcurve. In other words, the double-peaked lightcurve is due to the shape of the object, and it is clear that the asymmetry of the maxima is due to the spot/hemispheric albedo variation. So, the difference between both maxima (and both minima) gives us information about the albedo variation on the object's surface. In the case of Haumea, such a difference between both maxima is around 0.04~mag (or 4\%), and there is
also a difference between both minima of around 0.04~mag \citep{Lacerda2008}. Therefore, the total change can amount to 0.08~mag. Haumea is not the only TNO presenting such a characteristic. We can cite: i) Varuna with a difference around 0.1~mag (10\%) (\citet{Thirouin2010, Thirouin2013}, and Ortiz et al. In prep), ii) 2003~VS$_{2}$ also presents an asymmetric lightcurve with a 0.04~mag difference \citep{Thirouin2010, Sheppard2007, Ortiz2006}. The observed asymmetric lightcurves can be perfectly explained thanks to the lightcurve modeling of objects with spot or hemispheric albedo variations reported by \citet{Lacerda2008} \citep{Lellouch2010, Snodgrass2010, Carry2012, Lockwood2014}. Besides these cases, there is an even more well known case: Pluto. Pluto is a MacLaurin body whose lightcurve is exclusively due to albedo variations on its surface. In conclusion, based on the typical hemispherically averaged albedo, and the best Maxwellian fit distribution \citep{Thirouin2013}, we estimate that 0.15~mag is a good measure of the typical variability caused by albedo features. 
On the other hand, we know that lightcurve amplitude of large and small TNOs are significantly different \citep{Lacerda2003, Lacerda2006}. These authors, based on numerical and observations evidence, noticed a cut-off for objects with a diameter $\sim$400~km (with an albedo of 0.04). They demonstrated that large objects have nearly spherical shapes and small objects have irregular shapes. Based on a larger sample and using an albedo of 0.12, this cut-off is closer to $\sim$250-300~km \citep{Vilenius2012, Vilenius2014}. From observations and numerical simulations, we can conclude that objects with a diameter larger than 250-350~km (conservative cut-off) are spherical whereas smaller objects have an elongated shape \citep{Duffard2009, Lacerda2006, Lacerda2003, Leinhardt2000}. 
In case of asymmetric lightcurve, we always chose the double-peaked option because it clearly shows a complex shape and/or surface variations that cannot be explained with a single-peaked lightcurve.


\section{Photometric results}
\label{sec:photo}

In this section, we discuss our short-term variability results. We report new data for eleven objects. For one object we present a new analysis of previously published results by our team plus additional data we obtained. For the first time, rotational periods and lightcurve amplitude are reported for the entire family (except Haumea's satellites). To present a complete study, we also focus on several candidates. Observations of candidates are challenging because these objects are small and faint. Only a few attempts of short-term variability of faint objects have been published (e.g. \citet{TrillingBernstein2006, Kern2006_phd, Kern2006}). HST is the most prolific tool to study these faint objects, however, thanks to 4~m class telescope, we reached objects with a visual magnitude up to 24 (faintest objects observed under excellent weather and seeing conditions).    \\
 
Lomb periodograms and lightcurves for all objects are provided in Figure 1 to Figure 20. We plotted all lightcurves over two cycles (rotational phase from 0 to 2) for better visualization of the cyclical variation. For each lightcurve, a first or second order Fourier series is used to fit the photometric data. Error bars for the measurements are not shown on the plots for clarity but one-sigma error bars on the relative magnitudes are reported in the supplementary material (see Table~\ref{Tab:shortterm}). We must point out that when we combined several observing runs obtained at different epochs, light time correction of the data is required (see \citet{Thirouin2013} for more details about data combination). Our photometric results are reported in Table~\ref{Summary_photo}. A complete summary of the short-term variability of the Haumea family members and candidates can be found in Table~\ref{Tab:allphoto}.


\subsection{(24835) 1995~SM$_{55}$}

\citet{SheppardJewitt2003} observed this object for several nights on October and November 2001 with the University of Hawaii 2.2~m telescope. Based on the October data set, they reported a scattered photometry and no rotational period estimation. With the additional November data set, they suggested a single-peaked rotational period of 4.04~h or a double-peaked periodicity of 8.08~h and an average peak-to-peak amplitude of 0.19$\pm$0.05~mag (they only reported photometric amplitude estimated from apparent maximum and minimum, and not lightcurve amplitude obtained thanks to a lightcurve fit as it has been done in this work). Unfortunately, in both cases, the curves were too noisy given the photometric uncertainties. \citet{SheppardJewitt2003} concluded that the amplitude of the lightcurve may be variable from night to night. Such variations could be due to: i) the presence of a companion, ii) cometary activity, or iii) complex rotational state. \citet{SheppardJewitt2003} pointed out that this object has been investigated for binarity with the \textit{Hubble Space Telescope} and that no satellite with a separation $\geq$0.1$\arcsec$ and having a magnitude difference $\leq$2.5 was found. 1995~SM$_{55}$ is one of the bluest TNOs which could be attributed to recent excavation (due to a collision, for example) of its volatile-rich interior \citep{Hainaut2002}. On the other hand, the lightcurve amplitude may be due to freshly exposed material by cometary activity \citep{Hainaut2002}.  \\

\begin{figure}
\includegraphics[width=10cm, angle=180]{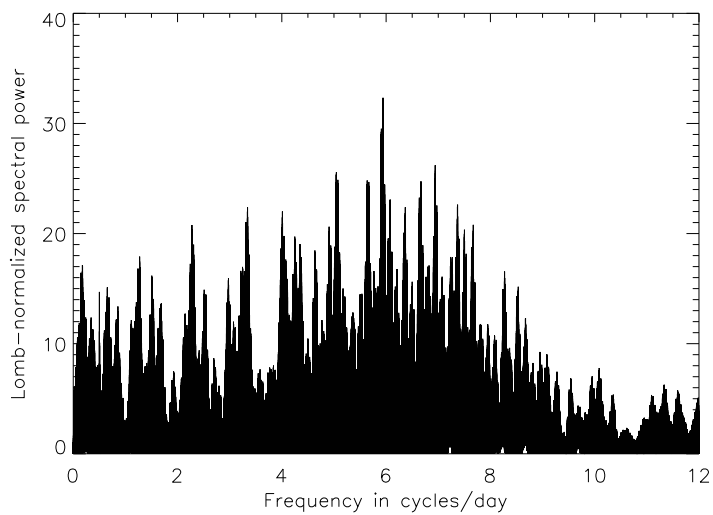}
\caption {\textit{Lomb-normalized spectral power versus frequency in cycles/day for 1995~SM$_{55}$}: the Lomb periodogram shows one main peak located at 4.04~h (5.94~cycles/day), and two aliases located at 4.77~h (5.04~cycles/day), and 3.45~h (6.94~cycles/day).}
\label{fig:Lomb_SM55}
\end{figure}

\begin{figure}
\includegraphics[width=10cm, angle=0]{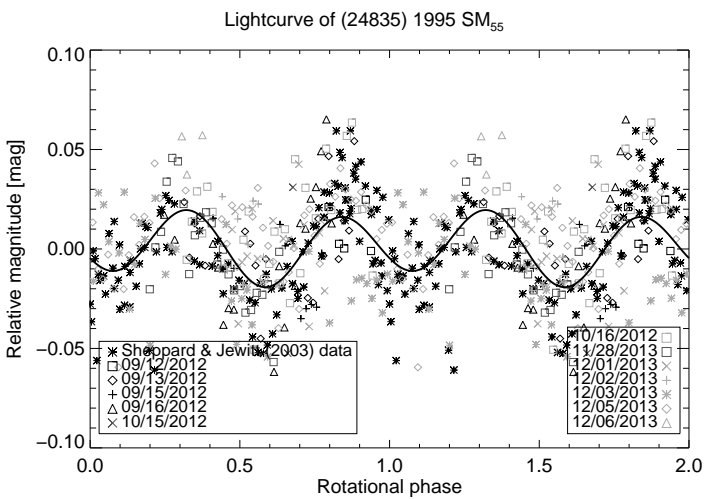}
\caption{\textit{1995~SM$_{55}$ lightcurve}: Rotational phase curve for 1995~SM$_{55}$ obtained by using a rotational period of 8.08~h. The continuous line is a Fourier fit of the photometric data. Different symbols correspond to different dates.}
\label{fig:LC_SM55}
\end{figure}

This object was observed in 2012, and 2013 to look for a possible change in the lightcurve. By merging our data with \citet{SheppardJewitt2003} data, the Lomb periodogram plotted in Figure~\ref{fig:Lomb_SM55} is obtained. The main peak is located at 5.94~cycles/day (4.04~h), and there are two other peaks with a lower spectral power located at 5.04~cycles/day, and at 6.94~cycles/day. The lightcurve is asymmetric with a first peak taller than the second one, and a second minimum deeper than the first one, so the double-peaked lightcurve with a rotational period of 8.08~h seems the best option. In Figure~\ref{fig:LC_SM55} is plotted the corresponding double-peaked lightcurve with a lightcurve amplitude of 0.04$\pm$0.02~mag. A second argument in favor of the double-peaked lightcurve is the goodness of the fit based on a reduced $\chi$$^2$ test ($\chi$$^2$=1.496 for the double-peaked option and $\chi$$^2$=1.510 for the single-peaked one). We also tested higher harmonics, but the $\chi$$^2$ test discarded all of them.  

In conclusion, we confirmed the rotational period obtained by \citet{SheppardJewitt2003}, as well as the lightcurve amplitude. We also report no significant change for this lightcurve over twelve years. Lightcurve reported by \citet{SheppardJewitt2003}, as well as our new lightcurve are noisy despite the high data quality. Possible explanations for such a lightcurve will be studied in a future work.


\subsection{(86047) 1999~OY$_{3}$}

\begin{figure}
\includegraphics[width=10cm, angle=180]{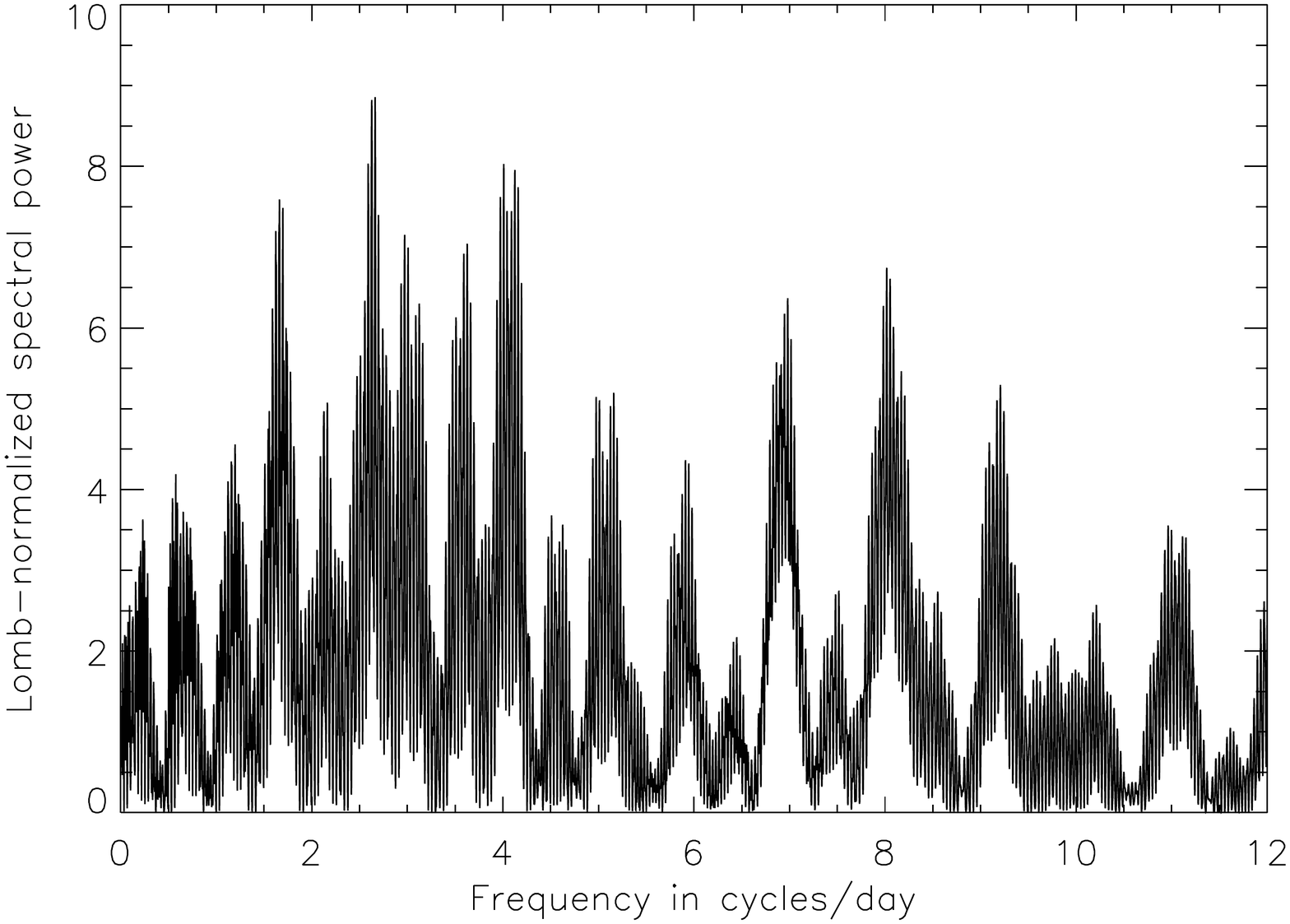}
\caption {\textit{Lomb-normalized spectral power versus frequency in cycles/day for 1999~OY$_{3}$}: the Lomb periodogram shows one main peak located at 9.01~h (2.66~cycles/day), and other peaks at 14.37~h (1.67~cycles/day), and 5.97~h (4.02~cycles/day).}
\label{fig:Lomb_OY3}
\end{figure}

\begin{figure}
\includegraphics[width=10cm, angle=0]{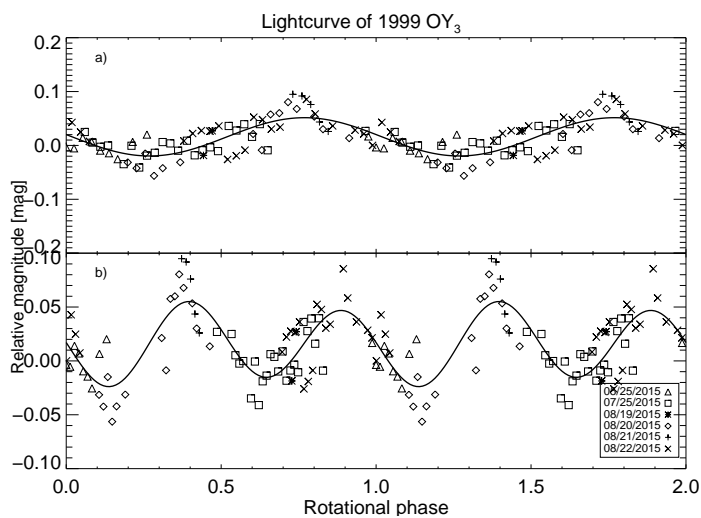}
\caption{\textit{1999~OY$_{3}$ lightcurve}: Rotational phase curve for 1999~OY$_{3}$ obtained by using a rotational period of 9.01~h (plot a)), and a double-peaked periodicity of 18.02~h (plot b)). Continuous lines are Fourier fit of the photometric data. Different symbols correspond to different dates. Same legend for both plots.}
\label{fig:LC_OY3}
\end{figure}

We report the first attempt of short-term variability study for this object. We observed 1999~OY$_{3}$ during several nights in 2015 with the DCT. We report two long observing blocks of about 6~h and several shorter blocks. Techniques used to derive the object rotation favored a periodicity around 9~h. Figure~\ref{fig:Lomb_OY3} showed one main peak at 2.66~cycles/day. Peaks around 4~cycles/day are consistent with the duration of our longer observing runs and so are not due to the object rotation. 
We obtained a single-peaked period of 9.01~h (Figure~\ref{fig:LC_OY3}, plot a)) and a double-peaked periodicity of 18.02~h (Figure~\ref{fig:LC_OY3}, plot b)). We favored the double-peaked option because minima/maxima are different by about 0.02~mag, and because the $\chi$$^2$ of the single-peaked fit is 1.633 whereas the double-peaked is 1.548. Considering the double-peaked option, we found a $\chi$$^2$=1.346 for the eighth harmonic. However, due to our data quality, we are not confident about this result. Only more data will confirm or not such a possible harmonic.


\subsection{(120178) 2003~OP$_{32}$}

\citet{Rabinowitz2008} presented 78 R-band observations of 2003~OP$_{32}$ obtained in 2006 with the 1.3~m SMARTS telescope. They proposed a single-peaked lightcurve with a periodicity of 4.845~h and an amplitude of 0.26~mag. \citet{Thirouin2010} also observed 2003~OP$_{32}$ during several runs between 2005 and 2007, and proposed a single-peaked lightcurve with a rotational period of 4.05~h and an amplitude peak-to-peak of 0.13$\pm$0.01~mag. \citet{Benecchi2013} observed 2003~OP$_{32}$, during 6 nights with the Ir\'{e}n\'{e}e du Pont 2.5~m telescope at Las Campanas Observatory (Chile). They favored a single-peaked rotational period of 4.85~h or a double-peaked rotational period of 9.71~h. Their peak-to-peak lightcurve amplitude is 0.18$\pm$0.01~mag.

\begin{figure}
\includegraphics[width=10cm, angle=180]{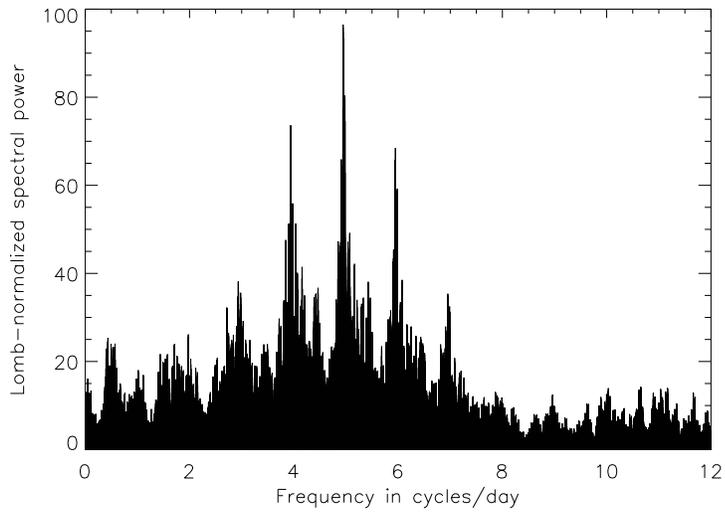}
\caption {\textit{Lomb-normalized spectral power versus frequency in cycles/day for 2003~OP$_{32}$}: the Lomb periodogram shows one main peak located at 4.85~h (4.95~cycles/day), and two aliases located at 6.07~h (3.96~cycles/day), and 4.03~h (5.96~cycles/day).}
\label{fig:Lomb_OP32}
\end{figure}

\begin{figure}
\includegraphics[width=10cm, angle=0]{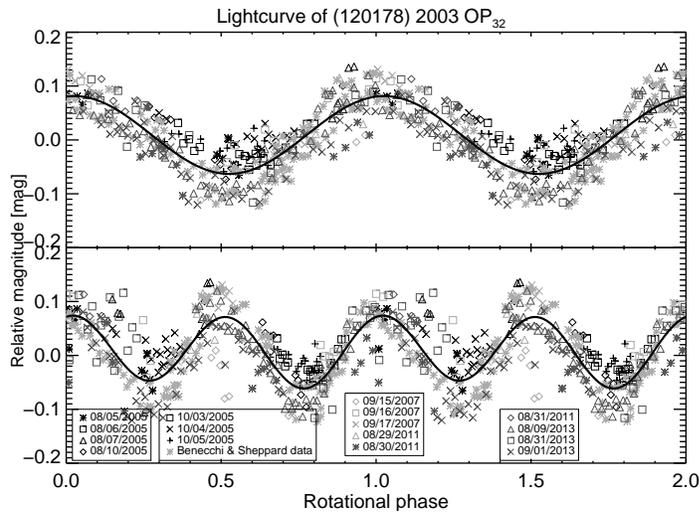}
\caption{\textit{2003~OP$_{32}$ lightcurve}: Rotational phase curve for 2003~OP$_{32}$ obtained by using a rotational period of 4.85~h (upper plot), and a rotational period of 9.71~h (double-peaked periodicity, lower plot). Continuous lines are Fourier fits of the photometric data. Different symbols correspond to different dates.}
\label{fig:LC_OP32}
\end{figure}

2003~OP$_{32}$ has been re-observed on 2011 and 2013 with the 2.2~m CAHA and the 1.5~m OSN telescopes. The Lomb periodogram of our 2005, 2007, 2011, 2013, and \citet{Benecchi2013} data sets altogether shows one peak located at 4.95~cycles/day (4.85~h) and two aliases located at 3.96~cycles/day (6.07~h) and at 5.96~cycles/day (4.03~h) (Figure~\ref{fig:Lomb_OP32}). All techniques confirm a periodic signature at 4.85~h with high spectral power. In Figure~\ref{fig:LC_OP32}, the corresponding single-peaked lightcurve with an amplitude of 0.14$\pm$0.02~mag is plotted. In conclusion, our and \citet{Benecchi2013} results completely ruled out the possibility of a large amplitude lightcurve noted by \citet{Rabinowitz2008}, and there is an agreement about the single-peaked periodicity of 4.85~h. 
Though we cannot rule out the double-peaked lightcurve, we favored the single-peaked option for this object for several reasons: i) the moderate lightcurve amplitude suggests albedo variation on the object's surface and not elongated shape behavior, ii) the lightcurve is symmetric, iii) because of the size of this object, it is more likely that it is a MacLaurin spheroidal object (see Section~\ref{sec:singledouble} for more details), and iv) $\chi$$^2$=1.559 for the single-peaked lightcurve and $\chi$$^2$=1.616 for the double-peaked.   \\


\subsection{2003~UZ$_{117}$}

Using the EMMI instrument installed at the New Technology Telescope of the European Southern Observatory (ESO), \citet{Perna2009} observed this object for $\sim$10.5~h during 2 nights on December 2007. As they mentioned, data points were not good enough to find an unambiguous rotational period. They suggested a rotational period of about 6~h. 

\begin{figure}
\includegraphics[width=10cm, angle=0]{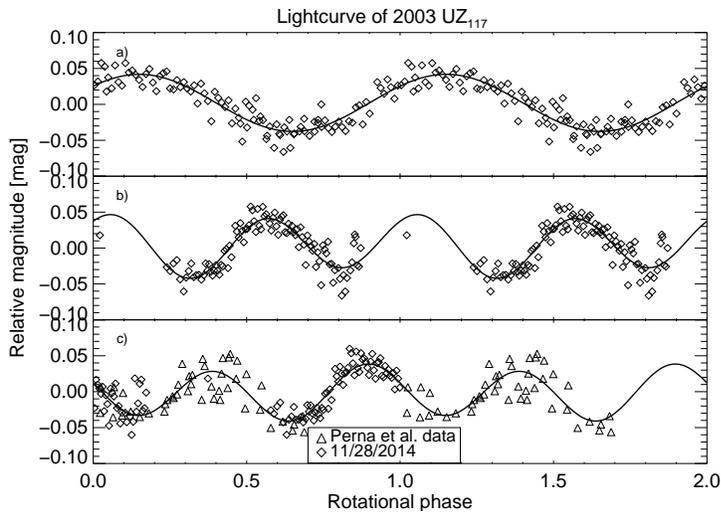}
\caption{\textit{2003~UZ$_{117}$ lightcurve}: Rotational phase curve for 2003~UZ$_{117}$ obtained by using a rotational period of 5.30~h (single-peaked lightcurve, plot a)), using a rotational period of 10.61~h (double-peaked lightcurve, plot b)), and using a rotational period of 11.29~h (double-peaked lightcurve, plot c)). The continuous lines are a Fourier fits of the photometric data. Different symbols correspond to different dates. Same legend for all the plots.}
\label{fig:LC_UZ117}
\end{figure}

\begin{figure}
\includegraphics[width=10cm, angle=180]{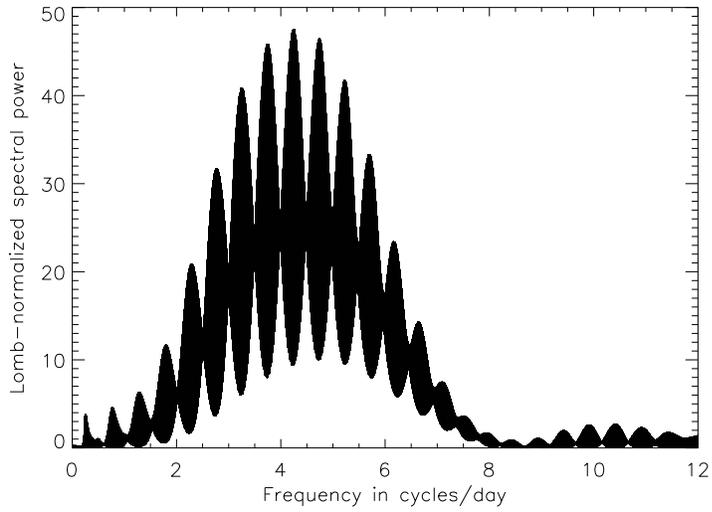}
\caption {\textit{Lomb-normalized spectral power versus frequency in cycles/day for 2003~UZ$_{117}$}: the Lomb periodogram shows one main peak located at 5.64~h (4.25~cycles/day) and two other peaks with a lower significance level at 6.42~h (3.74~cycles/day), and at 5.09~h (4.16~cycles/day).}
\label{fig:Lomb_UZ117}
\end{figure}

We observed 2003~UZ$_{117}$ during one night with the DCT on November 2014. Based on our DCT data, we obtained a single-peaked period of 5.30~h (Figure~\ref{fig:LC_UZ117}, plot a)) and a double-peaked periodicity of 10.61~h (Figure~\ref{fig:LC_UZ117}, plot b)). Using the ESO archive\footnote{Data can be downloaded at $\url{http://archive.eso.org}$}, we downloaded and re-reduced the images obtained by \citet{Perna2009}. By merging both data-sets reduced and analyzed the same way, we derived a double-peaked periodicity of 11.29~h (Figure~\ref{fig:Lomb_UZ117}). We favored the double-peaked option based on the fact that one of the minima is deeper than the other one and one of the maxima is taller than the other one ($\sim$0.01-0.02~mag). In Figure~\ref{fig:LC_UZ117} (plot c)) is plotted the corresponding lightcurve with an amplitude of 0.09$\pm$0.01~mag. We must point out that the rotational period of about 5.64~h was also an option based on our DCT data, but with a lower confidence level than the 5.30~h option. We calculated a $\chi$$^{2}$ of 1.246 for the second harmonic and 1.104 for the sixth harmonic. However, such a higher harmonic was only favored because \citet{Perna2009} data have a higher dispersion than our data. Therefore, we discarded this harmonic.


\subsection{(386723) 2009~YE$_{7}$}

\citet{Benecchi2013} observed 2009~YE$_{7}$ during 4 nights using the 2.5~m Ir\'{e}n\'{e}e du Pont telescope located at Las Campanas Observatory (Chile). They concluded that this object has a lightcurve amplitude $<0.2$~mag and they were not able to favor a rotational period based on their dataset. 

\begin{figure}
\includegraphics[width=10cm, angle=180]{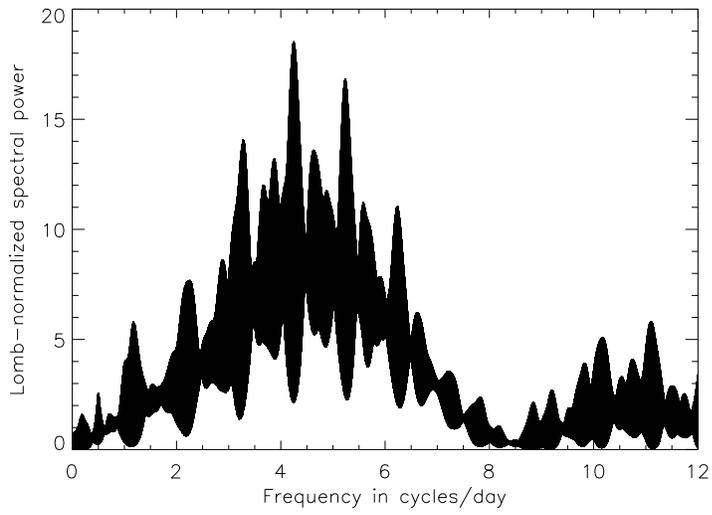}
\caption {\textit{Lomb-normalized spectral power versus frequency in cycles/day for 2009~YE$_{7}$}: the Lomb periodogram shows one main peak located at 5.65~h (4.25~cycles/day), and a second one located at  4.59~h (5.23~cycles/day).}
\label{fig:Lomb_YE7}
\end{figure}

\begin{figure}
\includegraphics[width=10cm, angle=0]{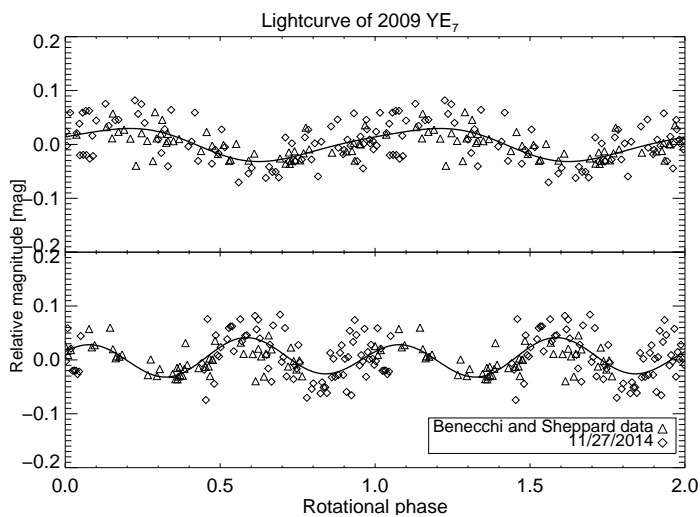}
\caption{\textit{2009~YE$_{7}$ lightcurve}: Rotational phase curve for 2009~YE$_{7}$ obtained by using a rotational period of 5.65~h (upper plot), and a rotational period of 11.30~h (double-peaked, lower plot). Continuous lines are Fourier fits of the photometric data. Different symbols correspond to different dates.}
\label{fig:LC_YE7}
\end{figure}

We observed this object during one night with the DCT in November 2014 under variable weather conditions. Based on our data-set, we derived a possible single-peaked lightcurve with a rotational period of about 5.5~h. By merging our sample and \citet{Benecchi2013} data, we obtained a single-peaked periodicity of 5.65~h (Figure~\ref{fig:Lomb_YE7}). The lightcurve amplitude is 0.06$\pm$0.02~mag (Figure~\ref{fig:LC_YE7}). Goodness of the fit is $\chi$$^{2}$=1.759 for the single-peaked lightcurve and $\chi$$^{2}$=2.005 for the double-peaked lightcurve. Based on the $\chi$$^{2}$ value the single-peaked option is favored. Though we cannot rule out the double-peaked lightcurve, the small lightcurve amplitude is compatible with albedo variation, and so we infer a single-peaked lightcurve \citep{Duffard2009, Thirouin2010}.  


\subsection{(315530) 2008~AP$_{129}$}

\begin{figure}
\includegraphics[width=10cm, angle=180]{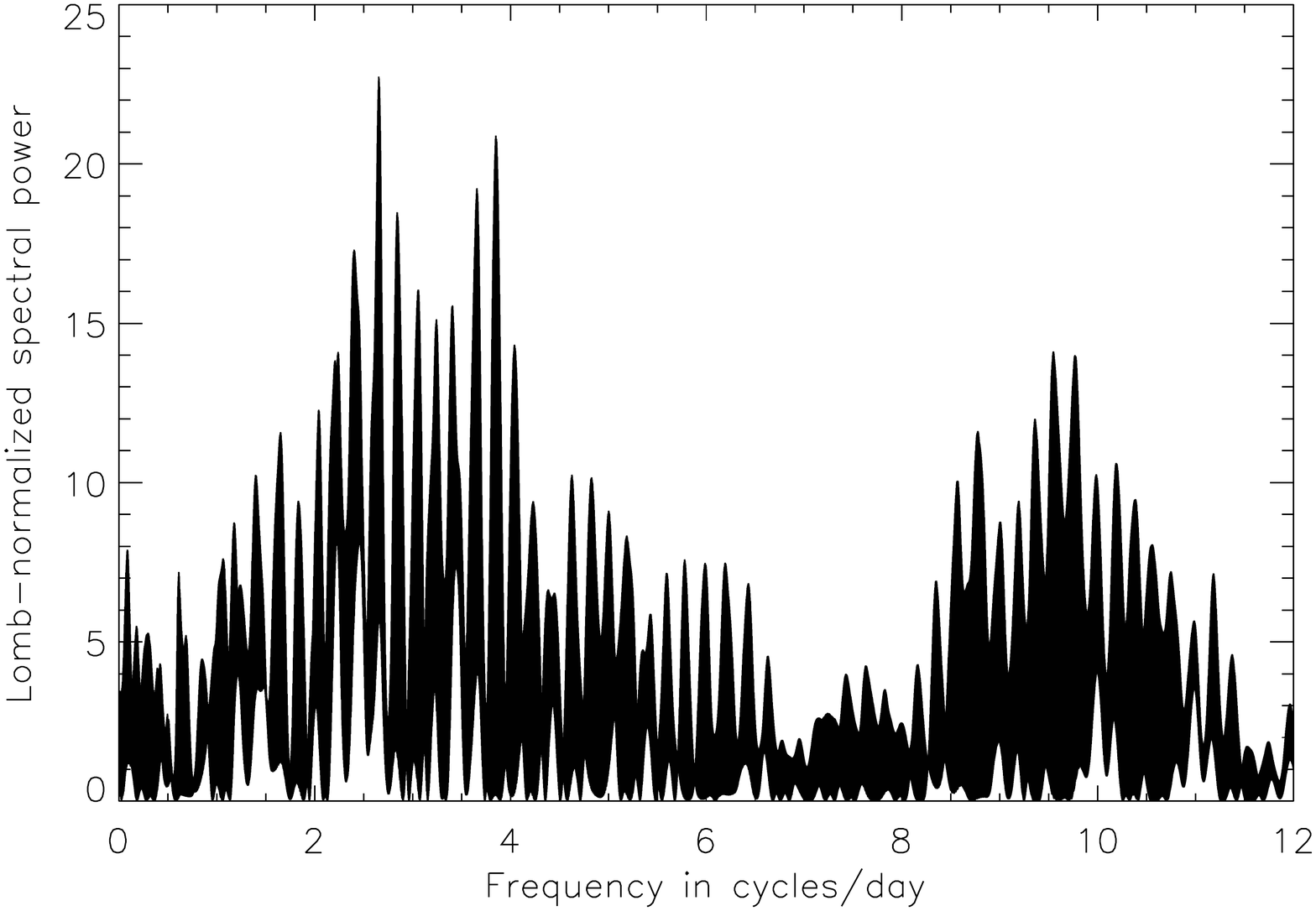}
\caption {\textit{Lomb-normalized spectral power versus frequency in cycles/day for 2008~AP$_{129}$}: the Lomb periodogram shows one main peak located at 9.04~h (2.65~cycles/day), and a second one located at 6.25~h (3.84~cycles/day).}
\label{fig:Lomb_AP129}
\end{figure}

\begin{figure}
\includegraphics[width=10cm, angle=0]{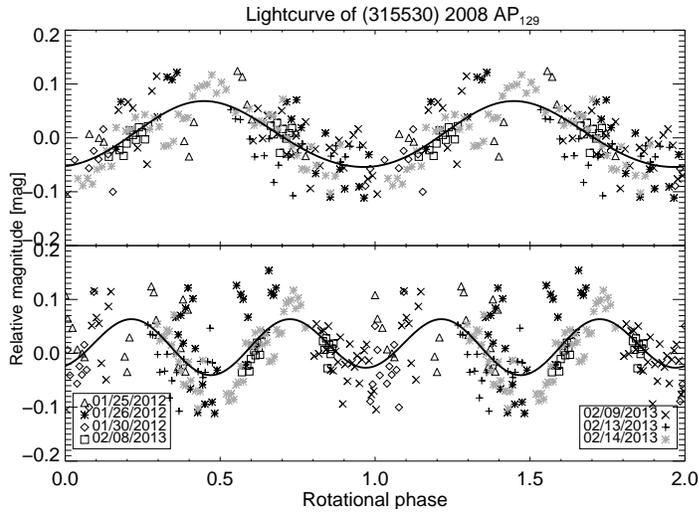}
\caption{\textit{2008~AP$_{129}$ lightcurve}: Rotational phase curve for 2008~AP$_{129}$ obtained by using a rotational period of 9.04~h (upper plot), and rotational period of 18.08~h (lower plot). The continuous line is a Fourier fit of the photometric data. Different symbols correspond to different dates.}
\label{fig:LC_AP129}
\end{figure}

2008~AP$_{129}$ did not previously have an observed lightcurve. We observed this object during a run in January 2012, in poor atmospheric conditions, and during two more runs in February 2013 with the 3.58~m TNG and the 1.5~m OSN telescope. The Lomb periodogram of our 2012 and 2013 data sets (Figure~\ref{fig:Lomb_AP129}) shows one main peak located at 9.04~h (2.65~cycles/day) and the second one with a lower spectral power is located at 3.84~cycles/day
(6.25~h). PDM, CLEAN, and Pravec-Harris techniques confirmed these two peaks with a higher
spectral power for the 9.04~h rotational period. In Figure~\ref{fig:LC_AP129}, the corresponding lightcurve using a rotational periodicity of 9.04~h is plotted. The amplitude of the curve is 0.12$\pm$0.02~mag. In
summary, 9.04~h is preferred but 6.25~h is also a possible period. In both cases, we preferred the single-peaked periodicity because of the small lightcurve amplitude, as well as the symmetric lightcurve and the potentially large size of this object. The goodness of the fit is $\chi$$^{2}$=1.539 for the single-peaked lightcurve and $\chi$$^{2}$=1.571 in the case of the double-peaked option. Therefore, the single-peaked lightcurve seems to be the best option, but the double-peaked option is not ruled out. Based on the data quality (data obtained under poor weather condition), we do not favor higher harmonics. Only more data will allow us to check if higher harmonics have to be considered.   


\subsection{1999~CD$_{158}$}

\begin{figure}
\includegraphics[width=10cm, angle=180]{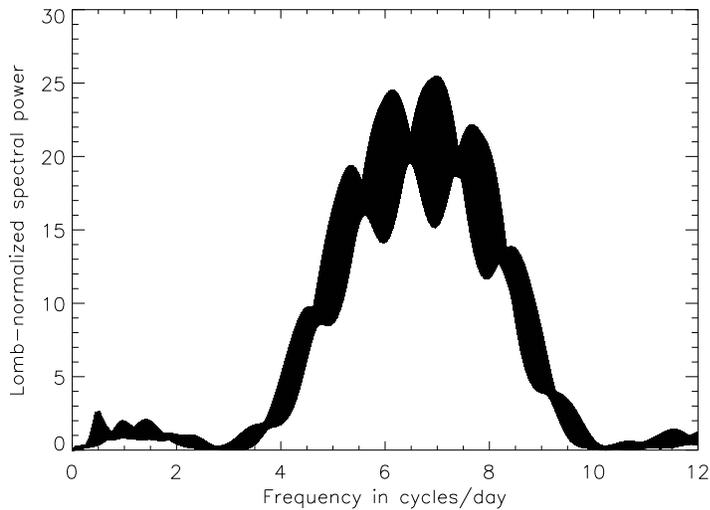}
\caption {\textit{Lomb-normalized spectral power versus frequency in cycles/day for 1999~CD$_{158}$}: the Lomb periodogram suggests a rotational period of 3.44~h (6.98~cycles/day).}
\label{fig:Lomb_CD158}
\end{figure}

\begin{figure}
\includegraphics[width=10cm, angle=0]{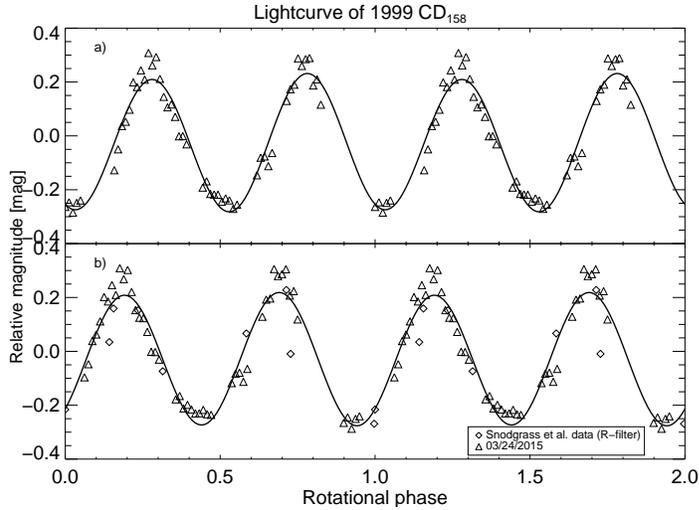}
\caption{\textit{1999~CD$_{158}$ lightcurve}: Rotational phase curve for 1999~CD$_{158}$ obtained by using a rotational period of 7.1~h (plot a)), and a rotational period of 6.88~h (plot b)) Continuous lines are Fourier fits of the photometric data. Different symbols correspond to different data-sets. }
\label{fig:LC_CD158}
\end{figure}

1999~CD$_{158}$ has not been previously observed for short-term variability. We observed this object during one night in March 2015 with the DCT. We derived a single-peaked period of 3.55~h and a double-peaked periodicity of 7.1~h. Lomb periodogram and other techniques confirmed this periodicity (Figure~\ref{fig:Lomb_CD158}). We favored the double-peaked option based on the large lightcurve amplitude. In Figure~\ref{fig:LC_CD158} (Plot a)) is plotted the corresponding lightcurve with an amplitude of 0.51$\pm$0.03~mag.  
\citet{Snodgrass2010} observed this object in 2008 using the 3.6~m New Technology telescope (NTT) located at La Silla Observatory (Chile). The main purpose of their work was to derive BVRI colors, and their data are not entirely suitable for lightcurve study (see \citet{Snodgrass2010} for more details and observing circumstances). Thanks to the ESO archive, we downloaded and re-reduced images obtained by \citet{Snodgrass2010}. By merging our sample and their R-data, we obtained a single-peaked periodicity of 3.44~h (Figure~\ref{fig:Lomb_CD158}). Based on the large lightcurve amplitude ($\Delta$m~=~0.49$\pm$0.03~mag) and the asymmetric peaks, we favored the double-peaked periodicity of 6.88~h (Figure~\ref{fig:LC_CD158}, plot b)). The best fit was obtain for the double-peaked lightcurve, and so our result is confirmed.  


\subsection{2000~CG$_{105}$}

\label{sec:CG105}

2000~CG$_{105}$  was observed only one night in 2015 with the DCT. From about 2~h of observations, we report a 0.2~mag amplitude variation. We searched for a rotational periodicity but, unfortunately, with only few observational hours, we are not able to propose a reliable rotational-period estimation. We try to merge our data with R-band images from \citet{Snodgrass2010}, but we are not able to derive a secure rotational period. However, based on \citet{Snodgrass2010} and our data, we can confirm the large variability of this object. 


\subsection{2002~GH$_{32}$}
 
\begin{figure}
\includegraphics[width=10cm, angle=180]{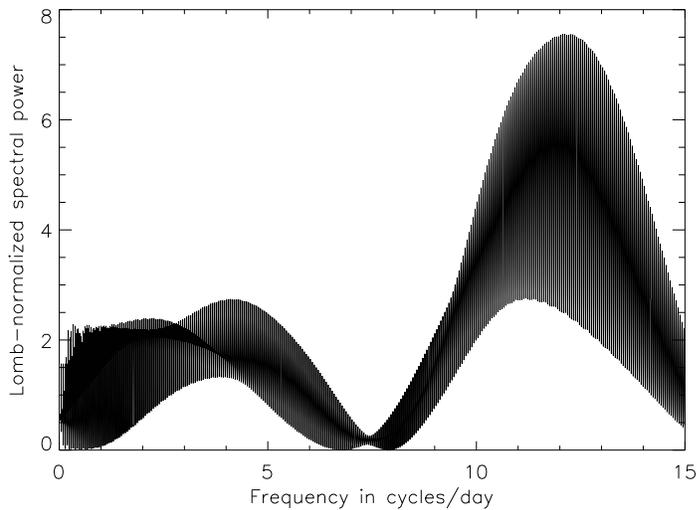}
\caption {\textit{Lomb-normalized spectral power versus frequency in cycles/day for 2002~GH$_{32}$}: the Lomb periodogram favors a periodicity of 1.99~h (12.06~cycles/day).}
\label{fig:Lomb_GH32}
\end{figure}

\begin{figure}
\includegraphics[width=10cm, angle=0]{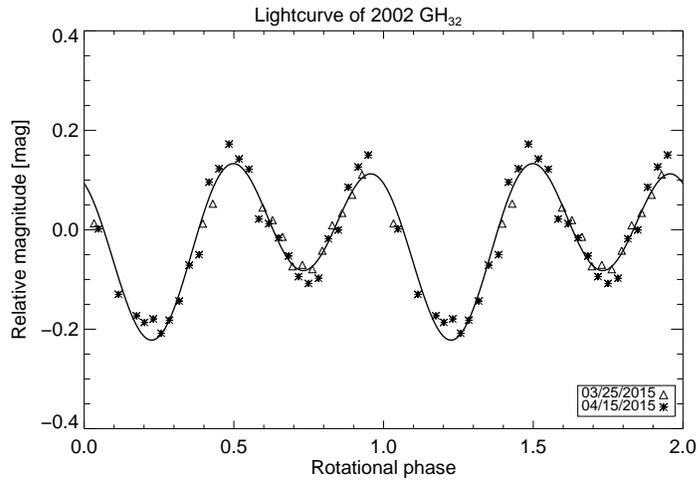}
\caption{\textit{2002~GH$_{32}$ lightcurve}: Rotational phase curve for 2002~GH$_{32}$ obtained by using a rotational period of 3.98~h. The continuous line is a Fourier fit of the photometric data. Different symbols correspond to different observing dates. }
\label{fig:LC_GH32}
\end{figure}

2002~GH$_{32}$ has not been observed for a lightcurve before. We observed this object during one night in March and one night in April 2015 with the DCT. We obtained a single-peaked period of 1.99~h. Lomb periodogram and other techniques confirmed such a periodicity (Figure~\ref{fig:Lomb_GH32}). However, the lightcurve is asymmetric and so, we favored a double-peaked periodicity of 3.98~h. In Figure~\ref{fig:LC_GH32} is plotted the corresponding lightcurve with a peak-to-peak amplitude of 0.36$\pm$0.02~mag. \citet{Harris2014} noticed that the fourth harmonic is potentially the primary harmonic for asteroid with a lightcurve dominated by shape and an amplitude up to $\sim$0.38~mag. Assuming the second harmonic, the goodness of the fit is $\chi$$^2$=1.310, whereas the fourth harmmonic has $\chi$$^2$=1.339. In conclusion, the second harmonic is favored. The first minimum of the curve is deeper than the second one by $\sim$0.08~mag. Such a lightcurve suggests that this object has a very elongated shape. In addition to similarities with Haumea, it is also interesting to point out that this object is the second fastest rotator in the Trans-neptunian belt after Haumea.


\subsection{2003~HA$_{57}$}

\begin{figure}
\includegraphics[width=10cm, angle=180]{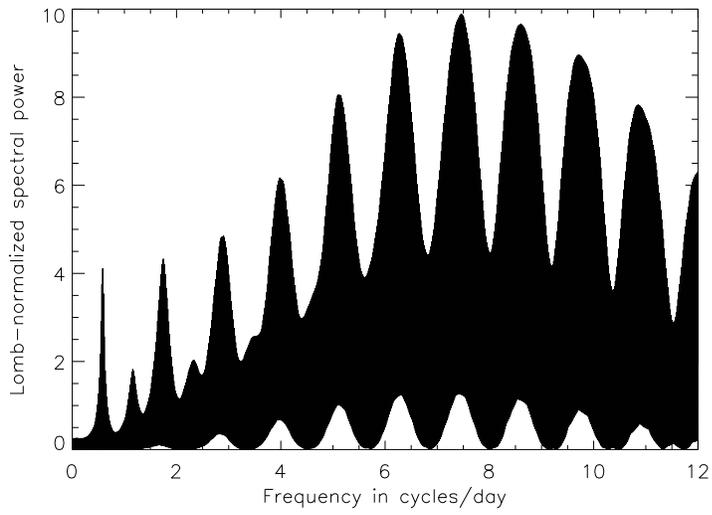}
\caption {\textit{Lomb-normalized spectral power versus frequency in cycles/day for 2003~HA$_{57}$}: the Lomb periodogram shows one main peak located at 3.22~h (7.46~cycles/day), and two other peaks at 3.83~h (6.27~cycles/day), and at 2.79~h (8.59~cycles/day).}
\label{fig:Lomb_HA57}
\end{figure}

\begin{figure}
\includegraphics[width=10cm, angle=0]{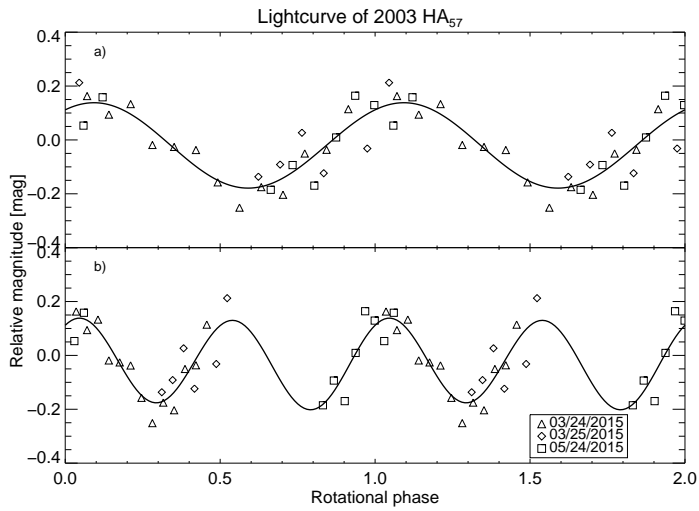}
\caption{\textit{2003~HA$_{57}$ lightcurve}: Rotational phase curve for 2003~HA$_{57}$ obtained by using a rotational period of 3.22~h (single-peaked lightcurve, plot a)), and a period of 6.44~h (double-peaked lightcurve, plot b)). Different symbols correspond to different dates. Continuous lines are a Fourier fit of the photometric data. Same legend for both plot. }
\label{fig:LC_HA57}
\end{figure}

We observed 2003~HA$_{57}$ during two consecutive nights in March 2015 and one night in May 2015 with the DCT. We derived a single-peaked lightcurve with a rotational period of 3.22~h (Figure~\ref{fig:Lomb_HA57}). All techniques confirmed such a periodicity. However, due to the large amplitude, such a curve is probably shape-dominated and so, we have to consider a double-peaked periodicity of 6.44~h. In Figure~\ref{fig:LC_HA57} are plotted the single- and double-peaked options (plot a) and b) respectively). We report an amplitude peak-to-peak of 0.31$\pm$0.03~mag. 
The $\chi$$^{2}$ value is 1.486 for the second harmonic. We also tested the goodness of the fit considering higher harmonics, but none of them are favored based on our data.


\subsection{2003~HX$_{56}$}

We observe this object $\sim$4~h during one night with DCT. With only few images we are not able to derive a secure rotational period. However, several constraints can be reported. We found a lightcurve amplitude higher than 0.4~mag and a rotational period higher than 5~h. Based on the large amplitude, we may have to consider a double-peaked lightcurve (rotational period higher than 10~h). 2003~HX$_{56}$ has not been observed for short-term variability before and so, more data are needed to complete this lightcurve.


\subsection{2005~GE$_{187}$}

\citet{Snodgrass2010} observed 2005~GE$_{187}$ in 2008 using the 3.6~m New Technology telescope (NTT) located at La Silla Observatory (Chile). Besides the fact that their main goal was to obtain BVRI colors, they noted that a single-peaked lightcurve with a period of 6.1~h and an amplitude of 0.5~mag seemed reasonably convincing for this object.  

\begin{figure}
\includegraphics[width=10cm, angle=180]{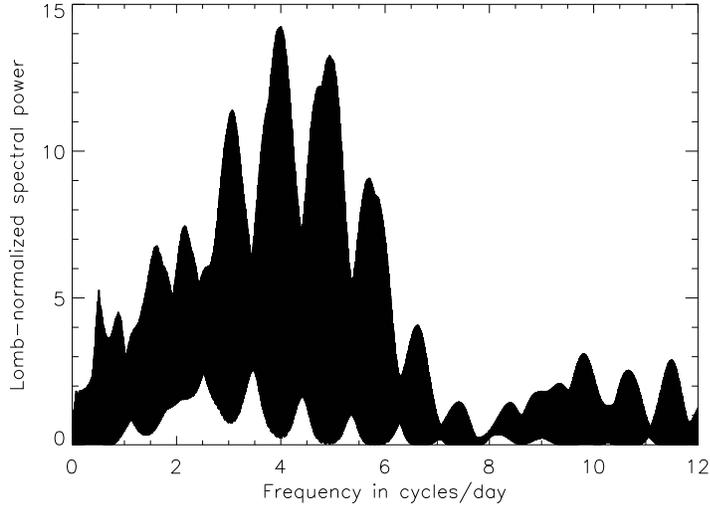}
\caption {\textit{Lomb-normalized spectral power versus frequency in cycles/day for 2005~GE$_{187}$}: the Lomb periodogram shows one main peak located at 5.99~h (4~cycles/day), and two other peaks at 7.81~h (3.07~cycles/day), and at 4.87~h (4.93~cycles/day).}
\label{fig:Lomb_GE187}
\end{figure}

\begin{figure}
\includegraphics[width=10cm, angle=0]{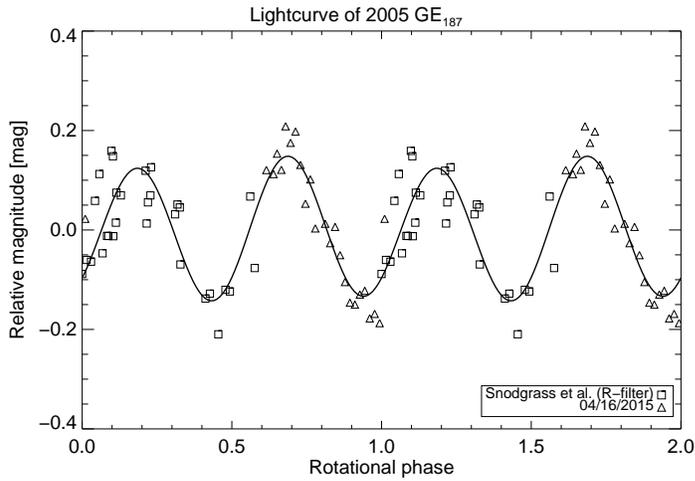}
\caption{\textit{2005~GE$_{187}$ lightcurve}: Rotational phase curve for 2005~GE$_{187}$ obtained by using a rotational period of 11.99~h. The continuous line is a Fourier fit of the photometric data. Different symbols correspond to different data-sets.}
\label{fig:LC_GE187}
\end{figure}

We observed this object during one night with the DCT on April 2015 under very poor seeing conditions. Based on our data-set, we derived a possible single-peaked lightcurve with a rotational period of 5.57~h and a lightcurve amplitude of 0.31~mag. We decided to take advantage of R-filter data from \citet{Snodgrass2010} and include them in our study. Using the ESO archive, we downloaded and re-reduced their images. By merging our sample and \citet{Snodgrass2010} R-data, we obtained a single-peaked periodicity of 5.99~h (Figure~\ref{fig:Lomb_GE187}). Based on the large lightcurve amplitude ($\Delta$m~=~0.29$\pm$0.02~mag) and the asymmetric peaks, such a curve is probably shape-dominated and therefore, we favored the double-peaked rotational period of 11.99~h (Figure~\ref{fig:LC_GE187}). For this object, we calculated a $\chi$$^{2}$ of 1.697 for the double-peaked lightcurve and $\chi$$^{2}$=1.959 for the single-peaked option. In conclusion, based on our data-set the best lightcurve is the double-peaked lightcurve. Higher harmonics have been considered but all of them have been discarded based on our data.    


\section{Rotational properties and derived parameters}

Here, we report lightcurves for all the family members (except Haumea's satellites). Despite studying the most complete sample to date, the Haumea family is only currently known to have eleven objects, therefore, results presented here are preliminary. 

\subsection{Rotational period distributions}

In the asteroid belt, members of dynamical families are thought to be fragments of collisions that could influence the spin properties of the family members \citep{Paolicchi2002}. If one assumes that the Haumea family is the result of a collision, one might expect the rotational properties of the family to be different from other TNOs. 

\begin{figure*}
\begin{center}
\includegraphics[width=15cm, angle=180]{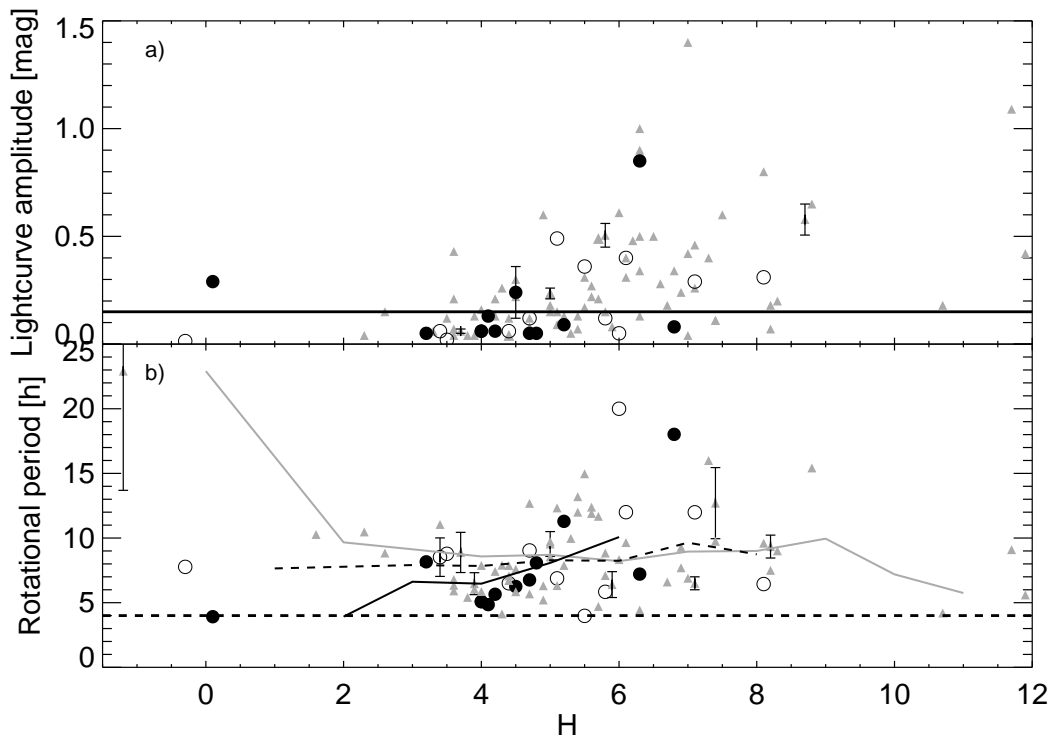}
 \caption{\textit{Lightcurve amplitude and rotational period versus absolute magnitude:} Black circles for the confirmed Haumea family members, open circles for the candidates, and gray triangle for the other TNOs. Same legends for both plots. \textit{Plot a)}: Continuous black horizontal line represents the shape- albedo-dominated lightcurve as in \citet{Thirouin2012, Thirouin2014}. \textit{Plot b)}: Spin barrier (dash horizontal line) around 4~h as suggested in \citet{Thirouin2010}. Absolute magnitudes from the Minor Planet Center (MPC). Lightcurve amplitudes and rotational periods are from Table~\ref{Tab:allphoto}, and \citet{Thirouin2013}. In case of multiple rotational periods proposed in the literature for the same object, we computed the mean period and the corresponding range of values. Pluto-Charon and Sila-Nunam are not plotted. Running means are also plotted, black continuous line for the Haumea family members, discontinuous black line for the candidates, and gray line for the other TNOs} 
\label{fig:multiplot}
\end{center}
\end{figure*}

In Figure~\ref{fig:multiplot} are plotted all the TNOs with a known rotational period and we highlighted the Haumea family members and candidates. A running mean is also reported for all samples considered. Other TNO and candidate samples exhibit a mostly flat running mean, therefore there is not a clear tendency between size and rotational period. But, smaller family members tend to spin slower than the larger members. \\

\begin{figure}
\includegraphics[width=15cm, angle=0]{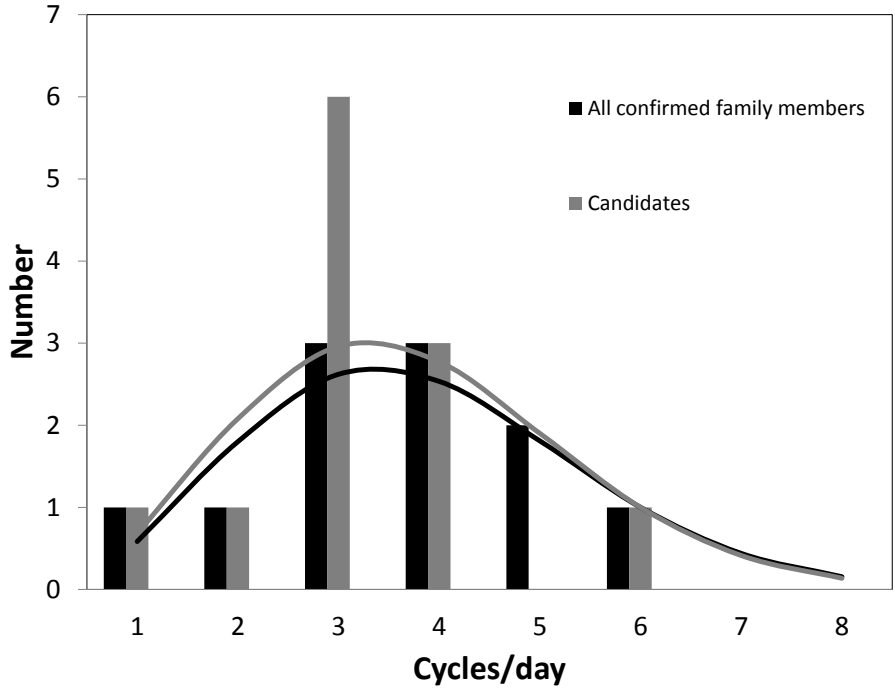}
 \caption{\textit{Number of objects versus rotational rate in cycles/day:} two different samples are plotted: confirmed members of the family (Number of objects (N)=11, black bars), and candidates (N=12, gray bars). A Maxwellian fit to the confirmed family members gives a mean rotational period of 6.27$\pm$1.19~h (continuous black line). The Maxwellian fit of the sample with the candidates gives a mean rotational period of 6.44$\pm$1.16~h (continuous gray line).   } 
\label{fig:Histo1}
\end{figure}

\begin{figure}
 \includegraphics[width=15cm, angle=0]{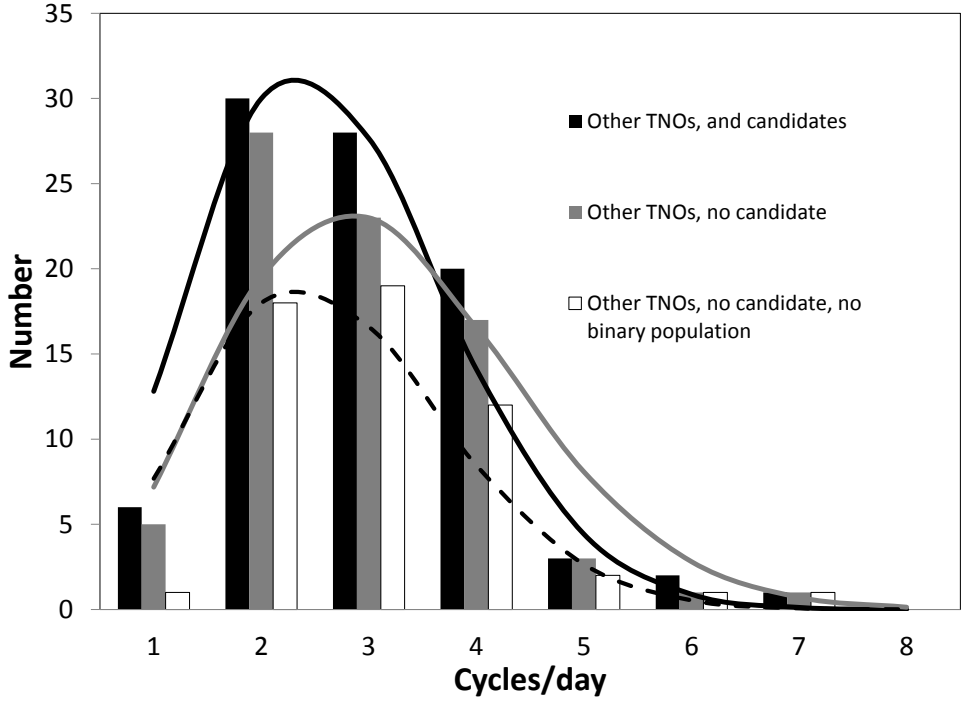}
  \caption{\textit{Number of objects versus rotational rate in cycles/day:} three different samples are plotted: other TNOs with candidates and 2008~AP$_{129}$ (i.e. all TNOs except confirmed members of the family) (number of objects (N)=90, black bars), other TNOs without candidates (N=78, gray bars), and other TNOs without the binary population, and candidates (N=53, white bars). A Maxwellian fit to the first sample gives a mean rotational period of 8.98$\pm$0.77~h (continuous black line). The Maxwellian fit of the second sample gives a mean rotational period of 7.65$\pm$0.54~h (continuous gray line). Maxwellian fit to the third sample gives a mean rotational period of 8.98$\pm$0.77~h (discontinuous black line) } 
 \label{fig:Histo2}
 \end{figure}

We noticed a tendency between size and rotational period for the Haumea family members suggesting that the smaller members of the family rotate slower than the biggest ones. With a significance level of 98.43$\%$, and the Spearman coefficient of 0.638, there is a strong evidence of correlation between absolute magnitude and rotational period for the family members (see Section~\ref{sec:corr}). We also tested the probability that the family members and other TNOs are from the same distribution using the 2D Kolmogorov-Smirnov test (KS test). The KS test estimates the maximum deviation between the cumulative distribution of both datasets to test the similarity (or not) between the two distributions (Df). Significance level of the KS test is a value between 0 and 1. Small values show that the cumulative distribution of the first dataset is significantly different from the second dataset. Considering two samples made of the family members and the other TNOs, we obtained a value of Df=0.54, and a significance level of 0.004, indicating that the rotational periods of the family are significantly different. In Figure~\ref{fig:Histo1}, rotational frequency of the family members and candidates are plotted (data from this work and the literature). The Maxwellian fits\footnote{We used Maxwellian fits based on \citet{Salo1987} and \citet{Binzel1989} works.} to the family sample gives a mean rotational period of 6.27$\pm$1.19~h, whereas the candidates have a mean rotational period of 6.57$\pm$1.14~h \footnote{Mean rotational rate in cycles/day, $\Omega$, is (8/$\pi$)$^{0.5}$ $\times$ $\sigma$ where $\sigma$ is the width of the distribution.}. In Figure~\ref{fig:Histo2} are plotted several samples without the family members and with/without the candidates, and the binary population \footnote{We have shown that the rotational properties of the binary population are different from the non-binary one and so, we removed the binary population in some samples \citet{Thirouin2014}.}. Maxwellian fit to the other TNOs sample gives a mean rotational period 8.98$\pm$0.77~h. In Table~\ref{Meanperiod} are reported the mean rotational periods from Maxwellian fits as well as average and median for several samples including/excluding the family members or the candidates. Standard error and standard deviation are indicated for all the samples. Without considering the error bars, average and median rotational periods of the family members indicate that the objects in relation with Haumea seem to rotate faster than the other TNOs. But, because of the large error bars regarding the family sample, such a tendency may not be true.   

Regarding the candidate sample, we found no evidence of correlation between absolute magnitude and rotational period. Running mean suggests a flat distribution similar to the one for the other TNOs sample. Maxwellian fit to the candidate sample favors a mean rotational period comparable to the family (but large error bars have to be considered). The candidate sample may be "contaminated" by icy members of the family yet to be identify, as well as objects without relation with the Haumea family. Despite the fact that 2008~AP$_{129}$ is not a confirmed member of the family, it follows the tendency between size and rotational period as the rest of the family.    \\ 

Several families have been identified and studied in the Main Belt of asteroids. Based on the Koronis and Eos family studies, the largest fragments of the families appear to have relatively similar rotation rates \citep{Binzel1989}. In other words, the largest fragments "remember" the spin rate of their parent body. The largest\footnote{Haumea family member sizes
have large uncertainty mainly because albedo is only known for two objects (see Table~\ref{tab:HaumeaMembers}), and so we use here the absolute magnitude to report the largest fragments.} members of the family are Haumea, 2002~TX$_{300}$, and 2003~OP$_{32}$ with rotational periods between $\sim$4~h and $\sim$8~h. The lightcurve of 2002~TX$_{300}$ is very flat and thus the rotational period may have significant uncertainty \citep{SheppardJewitt2003, Ortiz2004, Thirouin2010, Thirouin2012}. If 2003~OP$_{32}$ has a single-peaked lightcurve then two of the three largest members of the family, Haumea and 2003~OP$_{32}$ are also the fastest rotators of the family.
If the largest fragments of the family remember the spin rate of their parent body, we can conclude that the proto-Haumea was also a fast rotator with an elongated shape due to its rotation. We emphasize that a catastrophic collision is not able to create a fast-spinning elongated object. In fact, \citet{Leinhardt2010} simulated a catastrophic collision with exactly the same parameters proposed by \citet{Brown2007}. The results of their simulations show that the largest remnant is bigger than the current Haumea, and such a catastrophic collision produced a slow rotator with a rotational period of 28~h, far from the 3.92~h period of the current Haumea. Similar conclusion regarding the slow rotation has been obtained by \citet{Takeda2009}. Numerical simulations of catastrophic disruptions with enough resolution to resolve the shape of the largest remnant produce spherical objects, not fast-spinning elongated remnants \citep{Leinhardt2000, Leinhardt2009}. \\

Some experiments have focused on the rotation of fragments in catastrophic impacts for scenarios relevant to asteroid-like objects \citep{Fujiwara1981}. They performed catastrophic destruction of basalt targets by impacts of high-velocity projectiles. The rotational periods of the ejected fragments were measured as a function of particle size. Review of catastrophic disruption experiments using a wider range of materials has been reported by \citet{Martelli1994}. Both publications reported that the general tendency is that the smaller fragments have shorter rotational period than the bigger ones. Obviously, the size range of the fragments as well as the target composition are different for the Haumea case. But, if the relation noticed by \citet{Fujiwara1981, Martelli1994} is independent of size and composition, one could suggest that the Haumea family was not formed during a catastrophic collision. In conclusion, there is no clear explanation yet about such a relation between size and rotational period in the Haumea family. 

\subsection{Lightcurve amplitude distributions}

\begin{figure}
\includegraphics[width=10cm, angle=0]{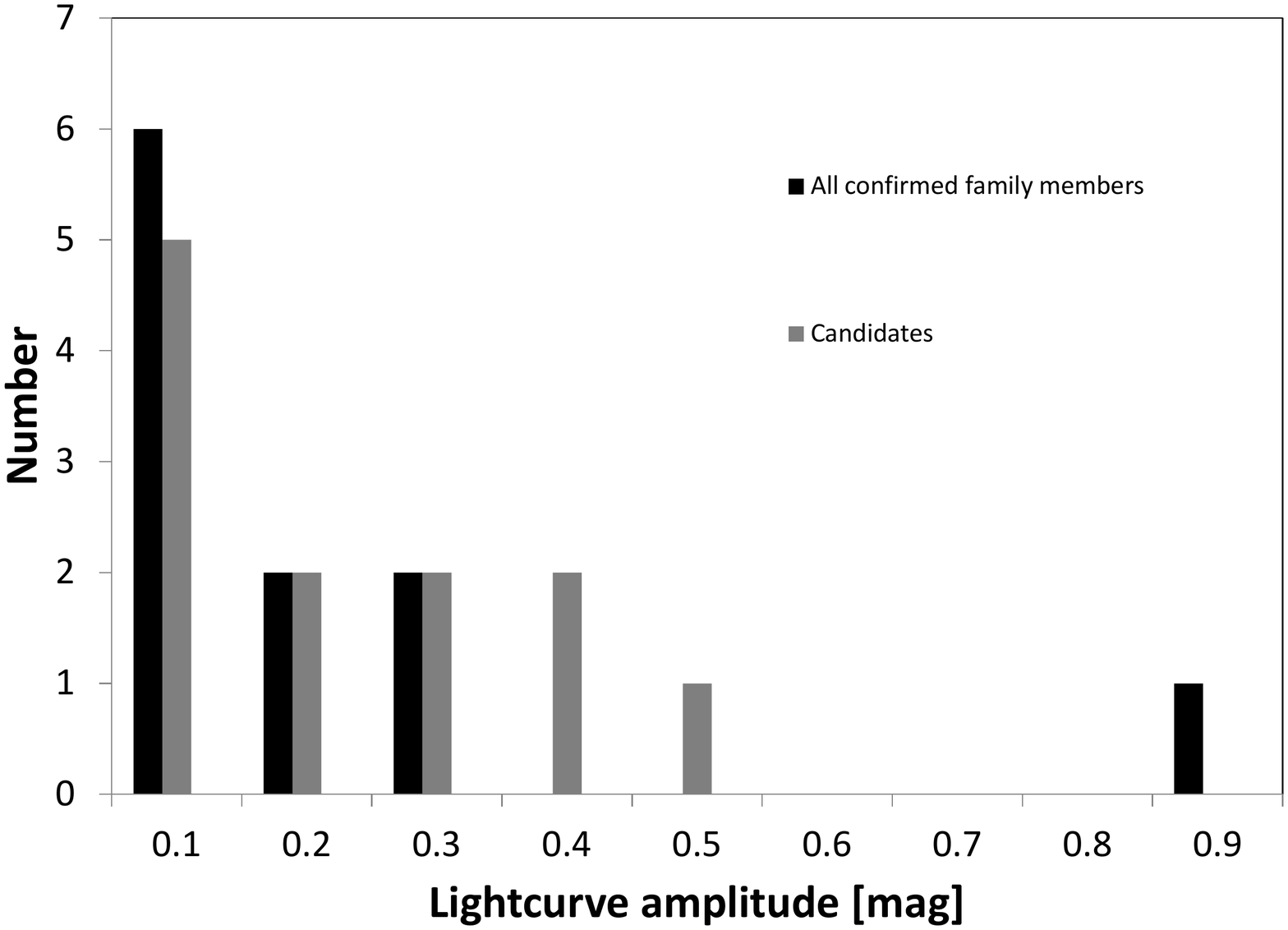}
\\
\includegraphics[width=10cm, angle=0]{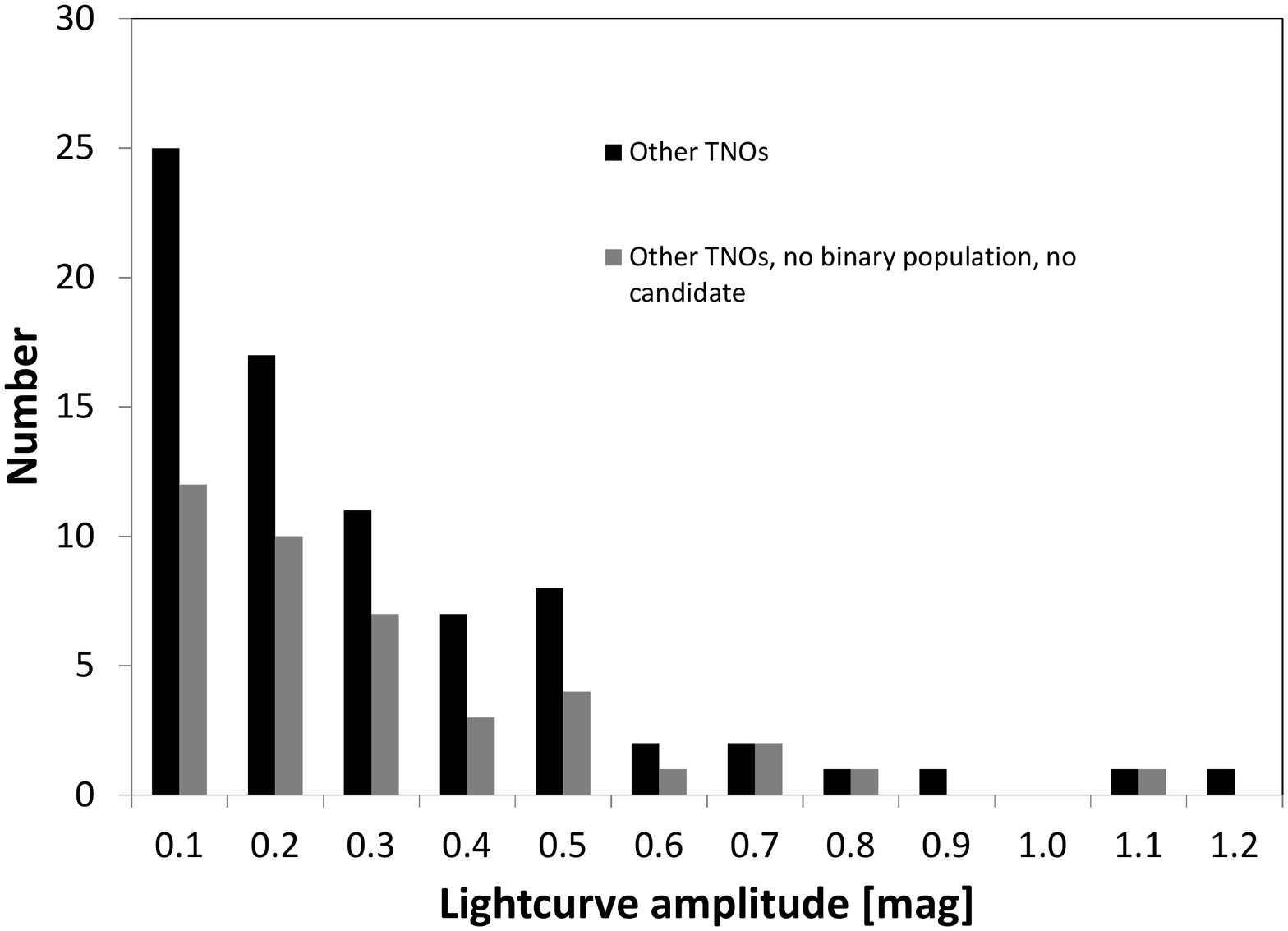}
 \caption{\textit{Number of objects versus lightcurve amplitude:} \textit{Upper plot}: two different samples are plotted; confirmed members of the family (number of objects (N)=11, black bars), and candidates (N=12, gray bars). \textit{Lower plot:} two different samples are plotted: other TNOs (i.e. all TNOs except confirmed members of the family) (number of objects (N)=90, black bars), other TNOs, except the binary population, the confirmed and candidates members of the family (N=54, gray bars). } 
\label{fig:histo_amplitud}
\end{figure}

In Figure~\ref{fig:histo_amplitud}, we focused on the lightcurve amplitude distributions for objects with/without relation with Haumea. Only two members of the family present a large lightcurve amplitude, Haumea and 2003~SQ$_{317}$, and maybe 1996~TO$_{66}$. The mean lightcurve amplitude for the confirmed members is 0.19~mag, and only 0.12~mag without the contact binary, 2003~SQ$_{317}$. So, most of the family members have a low lightcurve amplitude and thus should be considered as spherical (or nearly spherical) objects also known as MacLaurin spheroids. 
On the other hand, the candidates have larger lightcurve amplitude with a mean lightcurve amplitude of 0.20~mag. Such a mean value increases up to 0.28~mag if we only consider the smallest candidates with an absolute magnitude higher than 5. Objects with a large lightcurve amplitude have an elongated shape and are usually named Jacobi ellipsoids. This tendency has been already reported in the other TNOs sample \citep{Duffard2009, Thirouin2012, Thirouin2014}. In fact, smaller objects tend to have higher lightcurve amplitude (Figure~\ref{fig:multiplot}). Thus, it seems reasonable to expect that the candidates (i.e the smallest objects) have higher lightcurve amplitude. However, we cannot totally discard that such a tendency has some connection to the formation of the Haumea family (assuming that the candidates are members of the family).

\subsection{Size, and density}

\label{sec:densityestimate}

According to \citet{Binzel1989}, if we assume TNOs as triaxial ellipsoids with axes a$>$b$>$c and rotating along the c-axis, the lightcurve amplitude ($\Delta$${m}$) varies as a function of the observational (or viewing) angle $\xi$ as: 
\begin{equation}
\Delta m = 2.5 \log \left(\frac{a}{b}\right) - 1.25 \log \left(\frac{a^2 \cos ^2 \xi + c^2 \sin ^2 \xi}{b^2 \cos ^2 \xi + c^2 \sin ^2 \xi}\right)
\end{equation}
The lower limit for the object elongation (a/b) is obtained assuming an equatorial view ($\xi$=90$^\circ$). For a random distribution of spin vectors, the probability of viewing an object on the angle range [$\xi$, $\xi$+d$\xi$] is proportional to sin($\xi$)d$\xi$. The average viewing angle is $\xi$=60$^\circ$ \citep{Sheppard_phd}. We will consider viewing angles of 60$^\circ$ and 90$^\circ$.  

According to the study of \cite{Chandrasekhar1987} of equilibrium figures for fluid bodies, one can estimate lower limits for densities from rotational periods and the elongation of objects. We want to point out that it is difficult to estimate the size transition for an object to be in hydrostatic equilibrium or not. \citet{Lacerda2003, Lacerda2006} based on observational and numerical considerations, noticed a cut-off for objects with a diameter $\sim$400~km (assuming an albedo of 0.04). According to their studies, small objects have irregular shapes whereas large ones have a nearly spherical shape. This cut-off can be interpreted as a transition between spherical objects in hydrostatic equilibrium and rubble-pile objects. Lacerda and Luu works were based on a limited sample of lightcurve. By including newer results and assuming an albedo of 0.12 (more appropriate based on recent results), this cut-off is closer to $\sim$250-300~km \citep{Thirouin2013, Vilenius2012, Vilenius2014}. On the other hand, \citet{Duffard2009}, based on a simple Monte-Carlo model suggested that even small objects of about 120~km are in hydrostatic equilibrium. Finally, \citet{Lineweaver2010} derived the potato-to-sphere transition for icy body with a diameter of $\sim$200~km, and $\sim$300~km for rocky asteroids. This means that icy objects with a diameter smaller than about 200~km have an irregular shape whereas the biggest objects have a spherical shape. however, they are not considering that TNOs are rubble-piles. In this work, we used a conservative cut-off of 250-350~km for the transition between spheroidal and elongated objects. In conclusion, TNOs in the 200~km and below that size range can be rubble-piles. Collisional evolution models of the Transneptunian region indicate that objects in this size range have received several collisiones on average so they are likely re-accumulated objects. Besides, \citet{Duffard2009} indicated that the lightcurve statistics of small objects is consistent with the figures of hydrostatic equlibrium.  

\begin{figure}
\includegraphics[width=15cm, angle=0]{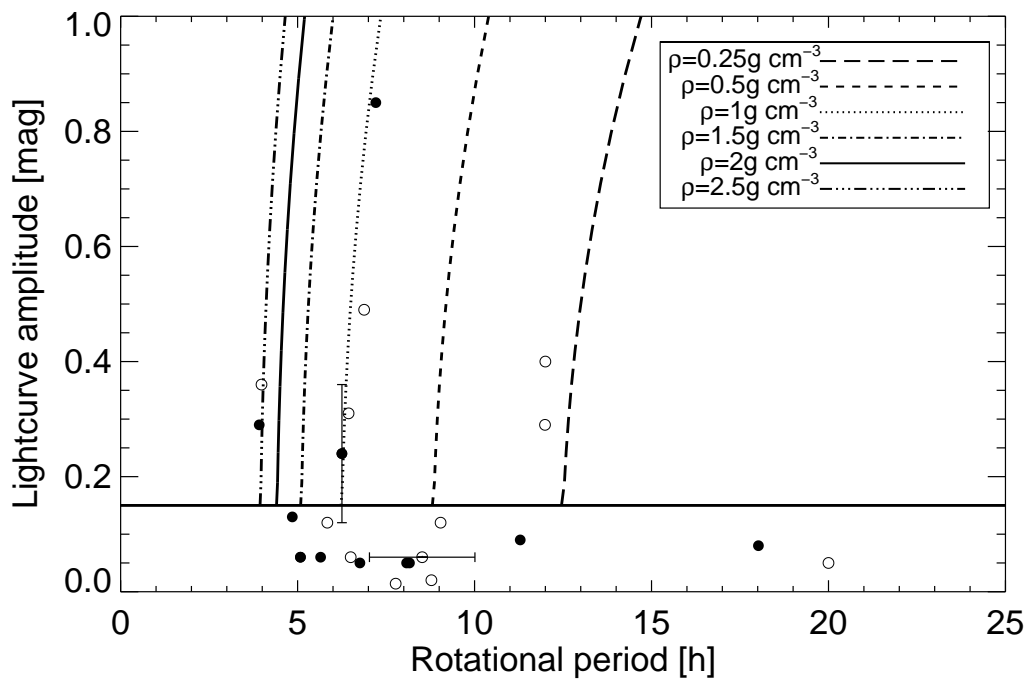}
\caption{\textit{Lightcurve amplitude versus Rotational period} for theoretical Jacobi ellipsoids of various densities compared with observations. Density values are indicated in the legend. Vertical lines have been computed using \cite{Chandrasekhar1987}. All objects presented in this work are shown: black circles for the confirmed Haumea family members, and open circles for the candidates. Continuous black horizontal line represents the shape- or albedo-dominated lightcurve as in \citet{Thirouin2012} and \citet{Thirouin2014}. Lightcurve amplitudes and rotational periods are from Table~\ref{Tab:allphoto}.   }
\label{fig:AmplPer}
\end{figure}

Assuming that a given TNO is a triaxial ellipsoid (Jacobi ellipsoid) in hydrostatic equilibrium, one can compute a lower density limit (Figure~\ref{fig:AmplPer} and Table~\ref{tab:Density}). Most of our targets have low amplitude lightcurves (i.e. low elongation), and their lightcurves are probably due to albedo effects. In other words, they are MacLaurin spheroids and the study on lower limit densities cannot be applied. In fact, most of the observed objects are far from the theoretical curves for acceptable values for the density which indicates that those objects are likely spheroids or are not in hydrostatic equilibrium \citep{Thirouin2014}. Only a few members have a high amplitude lightcurve ($>$0.15 mag): Haumea, 2003~SQ$_{317}$, and maybe 1996~TO$_{66}$. Haumea has a high density, $\sim$2.5~g cm$^{-3}$ \citep{Rabinowitz2006, Thirouin2010} whereas 1996~TO$_{66}$ seems to have a density higher than 1~g cm$^{-3}$ (Figure~\ref{fig:AmplPer}). 2003~SQ$_{317}$ is a contact binary with a density of 2.7~g cm$^{-3}$ (or 0.8~g cm$^{-3}$ assuming a single object, but based on the lightcurve, this object seems to be a contact binary and therefore the highest density will be considered) according to \citet{Lacerda2014}. All the candidates presented here have a large lightcurve amplitude and so can be considered as Jacobi objects. For the candidates, the range of densities varies from 0.9 to 2.6~g cm$^{-3}$, and varies from 0.12 to 1.7~g cm$^{-3}$ for the confirmed family members. Mean density for the members is 1.1~g cm$^{-3}$ (0.7~g cm$^{-3}$ without the contact binary, and Haumea), and is 1~g cm$^{-3}$ for the candidates. We want to emphasize the case of 2002~GH$_{32}$. This object is a very fast rotator with a double-peaked periodicity of 3.98~h and a lightcurve amplitude of about 0.36~mag. The morphology of the 2002~GH$_{32}$ lightcurve is similar to Haumea's. In fact, both objects present a fast rotation as well as an asymmetric lightcurve. One possible interpretation of such a lightcurve is that 2002~GH$_{32}$ has an elongated shape with strong albedo variation on its surface, as it is the case for Haumea \citep{Lacerda2008}. From the lightcurve, we derived a lower limit to the density of 2.6~g cm$^{-3}$ (equatorial view) or 2.7~g cm$^{-3}$ (viewing angle of 60$^{\circ}$). 

\begin{figure}
\includegraphics[width=12cm, angle=0]{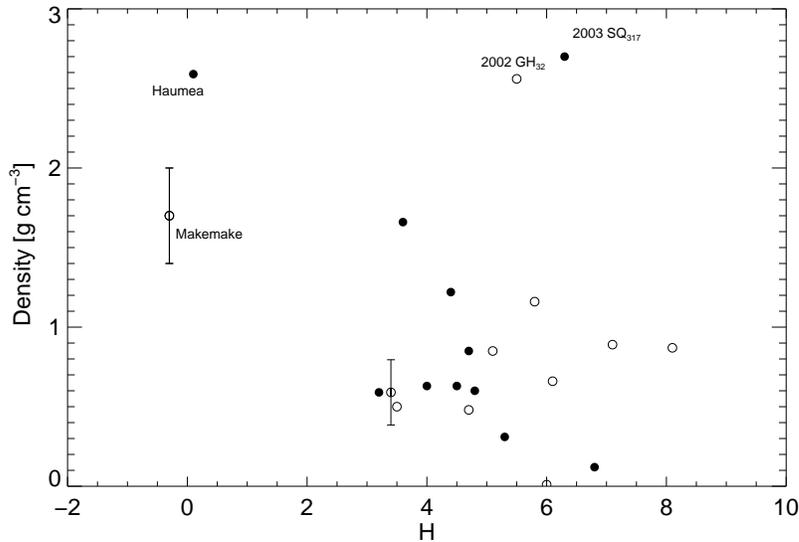}
\caption{\textit{Density versus absolute magnitude}: two samples are plotted: confirmed Haumea family members (black circles), and candidates (open circles). Most of the densities are only lower limit densities derived from lightcurves assuming an equatorial view (see Discussion). In the case of 2003~SQ$_{317}$, density assuming a contact binary and reported in \citet{Lacerda2014} has been used. It seems that the smallest members have the lowest density, except in the case of the contact binary which seems to have the highest density. Lower limits to density are reported in Table~\ref{tab:Density} for objects studied in this work. In the case of Makemake and Salacia, we used densities from \citet{Ortiz2012Makemake} and \citet{Stansberry2012}. For candidates not studied in this work but with a rotational period estimate, we derive their densities as in Section~\ref{sec:densityestimate}. 
Absolute magnitudes are from the Minor Planet Center (MPC).  
 }
\label{fig:DensityWater}
\end{figure}

In Figure~\ref{fig:DensityWater}, as previously noticed in \citet{Sheppard2008}, the biggest objects have higher densities than the smallest ones. Based on the family sample, an anti-correlation with $\rho$=-0.391, and a low significance level of 78.36$\%$ is found. Without the contact binary, 2003~SQ$_{317}$, a reasonably strong anti-correlation with $\rho$=-0.733 and a significance level of 97.33$\%$ is noticed. As pin-pointed by \citet{Lacerda2014}, the contact binary presents several atypical characteristics. To explain the lightcurve of this object, authors considered two cases: i) ellipsoidal object (Jacobi ellipsoid), and ii) compact binary near hydrostatic equilibrium (Roche binary). According to their models, 2003~SQ$_{317}$ may have a density between 670 and 1100~kg~cm$^{-3}$ considering a Jacobi object, or a density between 2050 and 3470~kg~cm$^{-3}$ based on a Roche model (see \citet{Lacerda2014} for more details). Assuming that the family members are from the icy mantle of the proto-Haumea, members would be expected to have a icy composition. The high density of the contact binary option suggests that it must be a fragment from the rocky core of the proto-Haumea. But, assuming an ellipsoidal object, the lower density indicates an icy composition. On the other hand, visible and near-infrared colours of this object confirm its membership, despite a much steeper phase function as the other members. In conclusion, as mentioned by \citet{Lacerda2014}, both options (Jacobi ellipsoid and Roche binary) are potential options. Only more observations over the next decade can confirm one of those options. Assuming a density of 860~kg~cm$^{-3}$ (i.e. Jacobi option) for 2003~SQ$_{317}$, we found an anti-correlation between density and absolute magnitude with a $\rho$=-0.636, and a significance level of 95.58$\%$ corresponding to a reasonably strong anti-correlation.         
Regarding only the candidate population, the trend between size and density is not evident ($\rho$=-0.251, significance level of 57.19$\%$).  \\

\subsection{Total Haumea family mass}

The mass of the family can be roughly estimated as follows. First, we computed the diameter and the mass of each confirmed member (Table~\ref{tab:HaumeaMembers}). The diameter (D) according to \citet{Pravec2007}, can be estimated by: 

 \begin{equation}
D = {\frac{K}{\sqrt{p} }} 10^{-0.2H}
 \end{equation}
where p is the geometric albedo, and H is the absolute magnitude. The constante K is: 
 \begin{equation}
K = 2 AU \times 10^{\frac{V_{sun}}{5}}
 \end{equation}
where V$_{sun}$ is the visual magnitude of the Sun. 
Assuming that the objects are spherical, the mass M is: 
\begin{equation}
M = {\frac{4}{3}} \pi \rho R^3
\end{equation}
where $\rho$ is the density and R is the radius of the object. 
By combining the previous equations, one can derive the mass, M, from: 

\begin{equation}
M = {\frac{\pi \rho}{6}} \left({\frac{K\times10^{-0.2H}} { \sqrt{p}}}\right)^3  
\end{equation}

An albedo of 0.7 has been assumed for the confirmed members of the family without a known albedo. Such an estimation is reasonable because all members are believed to be from the icy mantle of the proto-Haumea. For the candidates without a known albedo, we used albedos of 0.08 and 0.30 to derive a range of possible sizes and masses. Based on the masses computed and reported in Table~\ref{tab:HaumeaMembers}, we found a total mass of 4.06$\times$10$^{21}$~kg for the known family members (without Haumea, the total mass is 5.68$\times$10$^{19}$~kg~$\approx$~2$\%$M$_{Haumea}$ where M$_{Haumea}$ is the mass of Haumea). We did not include 2008~AP$_{129}$ as a member of the family because its membership is not confirmed yet (whose contribution would be very small). This mass estimation is obviously a lower limit because more small icy family members (and maybe rocky members) are expected to be found. On the other hand, as several members have no albedo reported, computed sizes indicated in Table~\ref{tab:HaumeaMembers} are only estimations. A more complete review will be proposed by Vilenius et al. (In prep) in which size and geometric albedos of several family members as well as some candidates will be derived thanks to thermal modeling of \textit{Herschel Space Observatory} data. \\

\subsection{Correlation/anti-correlation search}
\label{sec:corr}

We searched for correlations between physical (rotational period, and lightcurve amplitude) and orbital parameters. We used the Spearman rank correlation \citep{Spearman1904} because this method is less sensitive to atypical/wrong values and does not assume any population probability distribution. We computed the strength of the correlations by calculating the Spearman coefficient $\rho$ and the significance level (SL). The $\rho$ coefficient has values between -1 and 1. If $\rho$$>$0, there is a possible correlation, whereas $\rho$$<$0 indicated a possible anti-correlation and if $\rho$$=$0, there is no correlation. We consider a correlation as: i) strong if $|$$\rho$$|$$>$0.6, ii) weak if 0.3$<$$|$$\rho$$|$$<$0.6, and iii) non-existent if $|$$\rho$$|$$<$0.3.  
The significance of the $\rho$ parameter is measured by the SL: i) very strong evidence of correlation if SL$>$99$\%$ (i.e. 3~$\sigma$), ii) strong evidence of correlation if SL$>$97.5$\%$ (i.e. 2.5~$\sigma$), and iii) reasonably strong evidence of correlation if SL$>$95$\%$ (i.e. 2~$\sigma$). Such criteria have been used in several studies of correlations/anti-correlations between colors and orbital elements, for example in \citet{Peixinho2008, Peixinho2012, Peixinho2015}.   \\

Our search for correlations/anti-correlations is reported in Table~\ref{Tab:Correlations}. We used several samples, such as the entire family with/without the candidates, as well as other TNOs with/without the binary population (centaurs are not taken into account). As reported in \citet{Thirouin2014}, the binary population seems to exhibit distinct characteristics therefore special care has been taken to remove this population 
We also point out that the family and candidate samples are limited and so, care has to be taken regarding the correlation/anti-correlation detection and interpretation. Only correlations regarding the family and candidates with a significance level higher than 95$\%$ and $|$$\rho$$|$$\geq$0.3 are reported in Table~\ref{Tab:Correlations}.\\

\textit{Rotational period versus orbital elements}: \\
Correlations between rotational period and ascending node are only reported in several sub-samples but seem to be a specific characteristic of the candidates pool. It is also interesting to point out that such correlations are not presented in the other TNOs populations. We also report an anti-correlation between rotational period and inclination in the family.  
The most interesting and significant correlation is between rotational period and absolute magnitude. It seems that the biggest members of the family are rotating faster than the smallest ones. Such a correlation is only found in the family, not in the candidate nor other TNO pools. As mentioned in the previous section, this correlation, if real, is not understood yet. \\

\textit{Ligthcurve amplitude versus orbital elements}:  \\
A well known correlation is the one between lightcurve amplitude and absolute magnitude \citep{Duffard2009, Thirouin2010}. However, such a correlation is not found in the family sample. In fact, it seems that such a feature is only reported in the other TNOs, and candidates. Anti-correlations between lightcurve amplitude versus eccentricity and inclination are only reported in the other TNOs samples.


\section{Summary and Conclusions}

We have collected photometric data for several objects related to the Haumea family over the past five years using several facilities in Spain and United States of America. We present an homogeneous dataset composed of twelve objects: five family members, and seven candidates. We report rotational periods and lightcuve amplitude for most of them, but in two cases, we only report constraints. Half of studied objects have low lightcurve amplitudes (peak-to-peak amplitude less than 0.15~mag). Six of the twelve objects can be considered Jacobi ellipsoids with a high lightcurve amplitude due to the shape of the body. Some of the large amplitude lightcurves are asymmetric and such a fact can be explained by albedo variation(s) on the surface of the objects, as it is the case with the dark spot on Haumea's surface.  
We compared the rotational frequency distributions of the family members with/without candidates, and we conclude that the family members as well as candidates seem to rotate faster than the other TNOs (objects without relation with Haumea). We also have shown that the family members rotate at different rates according to their size. In fact, smaller members of the family rotate slower than the biggest fragments. Such a tendency is yet-to-be understood and will be discussed in future work. Rotational periods of the Haumea family members give us information about the characteristics of the Proto-Haumea. One of the conclusions, it that the Proto-Haumea was probably a fast rotator too and thus probably an elongated object (deformation due to the fast rotation).  

Regarding the lightcurve amplitude distribution, Haumea family members mostly show low amplitudes, besides the case of the contact binary 2003~SQ$_{317}$, and Haumea itself. However, it is interesting to point out that most of the candidates have moderate to large lightcurve amplitudes. Objects in relation with Haumea seem to follow the same tendency as objects without relation with Haumea, i.e. smaller objects have higher lightcurve amplitude thus have more irregular shapes.

From short-term variability study, we derive several physical properties such as axis ratio, and lower limit to the density. We have shown that the mean density for the members as well as for the candidates is around 1~g~cm$^{-3}$. In the family, bigger objects seem to have higher density, except the case of the contact binary whereas such a tendency is not evident in the candidate sample. 

The definition of the Haumea family is not clear. The classic definition considers that proper orbital elements and water ice detection are necessary to identify family members, but the enlarged definition suggests that rocky members without (or a small amount) water ice have to be considered too. Based on our study, it seems that the family members have rotational properties significantly different from the other TNOs. Therefore, such properties are probably useful to identify family members.

We also report the first lightcurve of 2002~GH$_{32}$. This object presents an asymmetric lightcurve and a very fast rotation. Our interpretation for such a lightcurve is that 2002~GH$_{32}$ is a very elongated object. 2002~GH$_{32}$ is the second fastest rotator in the trans-neptunian belt, after Haumea. We report a double-peaked periodicity of 3.98~h, and a lightcurve amplitude of 0.36~mag. Assuming this object as a triaxial ellipsoid in hydrostatic equilibrium, we derive a lower limit to the density of 2.6~g~cm$^{-3}$. Such a density is similar to Haumea's. We also want to emphasize the similarity of this lightcurve with Haumea's. In fact, both objects have similar rotational period and asymmetric lightcurve despite their size range difference.  

%
 
\acknowledgments

We thank an anonymous referee for her/his careful reading of the paper, and useful comments. 
We thank Will Grundy for very useful discussions. This research was based on data obtained at the Observatorio de Sierra Nevada which is operated by the Instituto de Astrof\'{i}sica de Andaluc\'{i}a, CSIC. Other results were obtained at the Telescopio Nazionale Galileo. The Telescopio Nazionale Galileo (TNG) is operated by the Fundaci\'{o}n Galileo Galilei of the Italian Istituto Nazionale di Astrofisica (INAF) on the island of La Palma in the Spanish Observatorio del Roque de los Muchachos of the Instituto de Astrof\'{i}sica de Canarias. Other results were obtained at the Isaac Newton Telescope (INT). The Isaac Newton Telescope is operated on the island of La Palma by the Isaac Newton Group (ING) in the Spanish Observatorio del Roque de Los Muchachos of the Instituto de Astrof\'{i}sica de Canarias (IAC). 
This research is also based on observations collected at the Centro Astron\'{o}mico Hispano Alem\'{a}n (CAHA) at Calar Alto, operated jointly by the Max-Planck Institut f\"{u}r Astronomie and the Instituto de Astrof\'{i}sica de Andaluc\'{i}a (IAA-CSIC). These results made use of Lowell Observatory's Discovery Channel Telescope (DCT). Lowell operates the DCT in partnership with Boston University, Northern Arizona University, the University of Maryland, and the University of Toledo. Partial support of the DCT was provided by Discovery Communications. LMI was built by Lowell Observatory using funds from the National Science Foundation (AST-1005313). We acknowledge the DCT operators: H. Larson, T. Pugh, and J. Sanborn; the OSN operators: F. Aceituno, V. Casanova, A. Sota. Special thanks to Heidi Larson who managed to fix all the technical problems we had during our DCT observing runs. We thank S. Levine, B. De Groff, L. Wasserman, P. Massey, D. Hunter, and D. Trilling for extra observing time at DCT. A. Thirouin was/is supported by AYA2008-06202-C03-01, NASA NEOO grant number NNX14AN82G, awarded to the Mission Accessible Near-Earth Object Survey (MANOS), and Lowell Observatory funding. J.L. Ortiz is supported by AYA2014-56637-C2-1-p which is a Spanish MICINN/MEC project, by the Proyecto de Excelencia de la Junta de Andaluc\'{i}a, J.A. 2012-FQM1776, and by FEDER funds.  \\

\clearpage

\begin{deluxetable}{lrrrrcrr}
\tablewidth{0pt}
\tablecaption{Observational circumstances$^{a}$. \label{Log_Obs}}
\tablehead{
\colhead{Object}           & \colhead{Date}      &
\colhead{Nb.}          & \colhead{r$_{h}$ [AU]}  &
\colhead{$\Delta$ [AU]}          & \colhead{$\alpha$ [$^{\circ}$]}    &
\colhead{Filter}  & \colhead{Tel.}  }
\startdata
(24835) 1995~SM$_{55}$ & 09/12/2012 & 9 & 38.437 & 37.892 & 1.27 & Clear & OSN\\
& 09/13/2012 & 20 & 38.436 & 37.878 & 1.26 & Clear & OSN\\
& 09/15/2012 & 23 & 38.436 & 37.851 &1.23 & Clear & OSN\\
& 09/16/2012 & 22 & 38.436 & 37.837 & 1.21 & Clear & OSN\\
& 10/15/2012 & 10 & 38.430 & 37.545 & 0.69 & Clear & OSN\\
& 10/16/2012 & 61 & 38.430 & 37.538 & 0.67 & Clear & OSN\\
 & 11/28/2013 & 25 & 38.350 & 37.441 & 0.58 & R & INT\\
& 12/01/2013 & 22 & 38.349 &37.459  &0.64  & Clear & OSN \\
& 12/02/2013 &13 & 38.349 &  37.466 & 0.66 & Clear & OSN \\
& 12/03/2013 & 52 & 38.349 &  37.473 &0.68  & Clear & OSN \\
& 12/05/2013 & 54 & 38.348 &37.487  & 0.72 & Clear & OSN \\
& 12/06/2013 & 6 & 38.348 & 37.495 &  0.74 & Clear & OSN \\
\hline
1999~CD$_{158}$ &  03/24/2015 &  51 & 46.288 & 46.933 &  0.93 & VR & DCT\\
\hline
(86047) 1999~OY$_{3}$ & 05/25/2015 &  5  &  41.127 & 41.264  &   1.40 & VR & DCT\\
& 07/25/2015 &  26  &  40.363 & 41.289  & 0.58 & VR & DCT\\
& 08/19/2015 &  4   & 40.304   & 41.300 & 0.25  & VR & DCT\\
& 08/20/2015 &  21   &  40.305 &  41.301  &  0.25 & VR & DCT\\
& 08/21/2015 &  5   &  40.307  &  41.301 &  0.26 & VR & DCT\\
& 08/22/2015 &  24   &   40.309  &   41.302 &  0.27 & VR & DCT\\
\hline
2000~CG$_{105}$ &  03/25/2015 & 9 &   46.220 & 47.025 & 0.72  & VR & DCT  \\
\hline
2002~GH$_{32}$ &  03/25/2015 &  12 &   43.619  &  42.929  & 0.95   & VR & DCT  \\
&  04/15/2015 &  30 &   43.624  &  42.713  &  0.56  & VR & DCT  \\
\hline
(120178) 2003~OP$_{32}$  & 08/29/2011 & 10 & 41.652 & 40.682  & 0.39  & KG1 &   CAHA\\
 & 08/30/2011 & 15 & 41.652 & 40.684 & 0.40 & KG1 &   CAHA  \\
 & 08/31/2011 & 5  & 41.652 & 40.866 & 0.41 & KG1 &  CAHA \\
 & 08/09/2013 & 52  & 41.842  & 40.900  &0.52   & Clear & OSN \\
 & 08/31/2013 & 33  &  41.848 &  40.873  & 0.36  & Clear & OSN  \\
 & 09/01/2013 & 47  & 41.848  & 40.875  &  0.37 & Clear & OSN  \\
\hline
2003~HA$_{57}$ &  03/24/2015 &  15  & 32.148 & 32.937 & 1.07 & VR & DCT\\
&  03/25/2015 &  7  &  32.138 &   32.937 & 1.06 & VR & DCT\\
&  05/24/2015 &   8 & 32.028  & 32.946  &  0.75 & VR & DCT\\
\hline
2003~HX$_{56}$ &05/26/2015 & 18   & 47.140  & 47.968  & 0.70 & VR & DCT\\
\hline
2003~UZ$_{117}$ &  11/28/2014 &109 & 38.944  & 38.005 &  0.45 & VR & DCT  \\
\hline
2005~GE$_{187}$ &  04/16/2015 &  26 &  28.887   & 28.106   & 1.26   & VR & DCT  \\
\hline
(315530) 2008~AP$_{129}$   & 01/25/2012 & 25 & 37.814 & 36.928 &  0.66 & Clear &  OSN\\
 & 01/26/2012 & 30 & 37.814 & 36.928& 0.66 & Clear &  OSN\\
 & 01/30/2012 & 15 & 37.815 & 36.929 & 0.66 & Clear &  OSN\\
 & 02/08/2013 &  20 & 37.930  & 37.051  &  0.69  & r' &  TNG \\
 & 02/09/2013 & 37 & 37.930  & 37.054  &  0.69  & Clear &  OSN\\
 & 02/13/2013 & 27 & 37.932  &37.068  &   0.73   & Clear &  OSN\\
 & 02/14/2013 & 63 &  37.932  & 37.073   &  0.74 & Clear &  OSN\\
 \hline
 (386723) 2009~YE$_{7}$   & 11/27/2014 & 90  & 50.694 & 49.832 & 0.55 &VR  & DCT\\
\enddata
\tablenotetext{a}{UT-Dates, heliocentric (r$_{h}$), and geocentric ($\Delta$) distances and phase angle ($\alpha$) of the observations are reported. We also indicate the number of images (Nb.) obtained each night, the filter used and the telescope (Tel.) for each observational run. "DCT" stands for the Discovery Channel Telescope, "OSN" stands for the Observatory of Sierra Nevada Telescope, "TNG" stands for the Telescopio Nazionale Galileo, "INT" for Isaac Newton Telescope, and "CAHA" stands for the Centro Astron\'{o}mico Hispano Alem\'{a}n telescope. Some data from others publications have been used, observational circumstances can be find in respective publications (see Section~\ref{sec:photo} for more details). }
\end{deluxetable}

\begin{scriptsize}
\begin{table}
\caption{\label{Tab:shortterm} The time series photometry of all the objects is provided in the “Center of astronomical Data of Strasbourg” (CDS). We present our photometric results: the name of the object and for each image we specify the Julian Date (JD, not corrected for light time), the relative magnitude (mag in magnitudes) and the 1-$\sigma$ error associated (Err. in magnitude), the filter (Fil.) used during observational runs, the phase angle ($\alpha$, in degree), topocentric (r$_{h}$) and heliocentric ($\Delta$) distances (both distances in Astronomical Units, AU). "Cle" stands for Clear filter. } 
 
\begin{tabular}{lccccccc} 
\hline
Object  & JD  & mag.   & Err.  &   Fil. & $\alpha$   & r$_\mathrm{h}$  & $\Delta$    \\
  &  [2450000+] & [mag] &  [mag]  &     &  [$^{\circ}$] &  [AU] &  [AU]   \\
\hline
\hline
315530 &	 	&	 	& 	& 	&	 &	 	&	  	\\
&	5952.44390	&	0.007	&	0.022	&	Cle	&	0.66	&	36.928	&	37.813	\\		
&	5952.45564	&	0.000	&	0.017	&	Cle  &	0.66	&	36.928	&	37.813	\\		
&	5952.46149	&	-0.007	&	0.015	&	Cle	&	0.66	&	36.928	&	37.813	\\		
&	5952.55931	&	-0.006	&	0.030	&	Cle	&	0.66	&	36.928	&	37.813	\\		
&	5952.56510	&	-0.034	&	0.015	&	Cle	&	0.66	&	36.928	&	37.813	\\
 
\hline
\hline

\end{tabular}
 
\end{table}
 
\end{scriptsize}

\begin{deluxetable}{ccccccc}
\tabletypesize{\scriptsize}
\rotate
\tablewidth{0pt}
\tablecaption{\label{Tab:allphoto} 
We listed the short-term variability of the confirmed Haumea family members, and the candidates. In case of multiple rotational periods, the preferred rotational period, according to the the authors of each study, is indicated in bold. Absolute magnitudes (Abs mag.) are from the Minor Planet Center (MPC) database. The complete reference list can be found after the table. Dynamical classification (Dyn. class) from \citet{Gladman2008} is also indicated.   }
\tablehead{
\colhead{Object}           & \colhead{Dyn. class }      &
\colhead{Single peak periodicity [h]}          & \colhead{Double peak periodicity [h]}  &
\colhead{Amplitude [mag]}          & \colhead{Abs mag.}    &
\colhead{Ref.}}
\startdata
\textit{Confirmed members}   &  &  &   &   &  \\
(136108) Haumea & Resonant &   - &  3.9154$\pm$0.0002  &   0.28$\pm$0.02   &   0.1  &  R06 \\
 & ... &  - &   3.9155$\pm$0.0001  &  0.29$\pm$0.02  &   ...  &  L08 \\
 & ... &   - &  3.92 & 0.28$\pm$0.02  &   ...  &  T10 \\
Namaka$^{a}$ & Resonant&   - &  -  &  - &   4.5   & - \\
Hi'iaka$^{a}$ & Resonant&   - &  -  &  - &  2.9   & - \\
(24835) 1995~SM$_{55}$$^{*}$ & Classical &   4.04$\pm$0.03  &  8.08$\pm$0.03  & 0.19$\pm$0.05  &  4.8  &  SJ03  \\
   & ... &   &8.08  & 0.05$\pm$0.02  &  ...   & TW   \\
(19308) 1996~TO$_{66}$$^{*}$ & Classical & \textbf{3.96$\pm$0.04} or 4.80$\pm$0.05 &  \textbf{7.92$\pm$0.04} or 5.90$\pm$0.05 or 9.6$\pm$0.1 &  0.26$\pm$0.03 &  4.5 &  SJ03\\
  & ... &  - &  11.9 &  0.25$\pm$0.05 &  ... &  B06\\
 & ... &  - &  6.25$\pm$0.03 &  0.12 to 0.33 &  ... &  H00\\
 & ... &  - &  - &  $<$0.10 &  ... &  RT99\\
(86047) 1999~OY$_{3}$$^{*}$ & Resonant &- &   18.02  &  0.08 & 6.8   &   TW \\
(55636) 2002~TX$_{300}$ & Classical & (\textbf{8.12} or 12.101)$\pm$0.08 & (\textbf{16.24} or 24.20)$\pm$0.08  &  0.08$\pm$0.02 &  3.2 &  SJ03\\
& ... &  7.89$\pm$0.03 &  15.78 &  0.09$\pm$0.08 &  ... &  O04\\
& ... &  - &  8.16 &  0.04$\pm$0.01 &  ... &  T10\\
& ... &  - &  8.15 or 11.7 &  (0.01 or 0.05)$\pm$0.01 &  ... &  T12\\
& ... &  - &  18.35 &  - &  ... &  S12\\  
(120178) 2003~OP$_{32}$$^{*}$ & Classical &  4.845$\pm$0.003 &  - &  0.26$\pm$0.04 &  3.6 &  R08\\
& ... &  4.05 &  - &  0.13$\pm$0.01 &  ... &  T10\\
& ... &  \textbf{4.85}/6.09 &  \textbf{9.71}/12.18 &  0.18$\pm$0.01 &  ... &  BS13\\
& ... &  4.85 &  - &  0.14$\pm$0.02 &  ... &   TW \\
2003~SQ$_{317}$$^{b}$ & Classical &  3.74    & - & 1  &  6.3  & S10 \\
& ... &  -    & 7.210$\pm$0.001 & 0.85$\pm$0.05  & ... & L14 \\
2003~UZ$_{117}$$^{*}$ & Classical &  $\sim$6  &  - & -  &  5.3 &  P09\\
           & ...          &  - &  11.29 & 0.09$\pm$0.01  & ... &  TW \\
(308193) 2005~CB$_{79}$ & Classical &  6.76 &  - &  0.05$\pm$0.02 &  4.7 &  T10\\
(145453) 2005~RR$_{43}$ & Classical &  - &  5.08$\pm$0.04 &  0.12$\pm$0.03 &  4.0 &  P09\\
& ... &  7.87 &  - &  0.06$\pm$0.01 &  ... &  T10\\
& ... &  - &  - &  $<$0.06  &  ... &  BS13\\
2009~YE$_{7}$ & Classical & - &  - & $<0$.20 &  4.4 &   BS13 \\
            & ...     &  5.65 & -& 0.06$\pm$0.02 & ... &  TW \\ 
\hline
\textit{Candidates} &    &  &   &   &  \\
(120347) Salacia & Classical  & - &   $\sim$17.5 &  0.2  &   4.4  &  S10 \\
& ... &      6.09 or 8.10  & - &   0.03$\pm$0.01  &   ...  & T10 \\
& ... &    - & - &   $<$0.04 &  ... &  B13 \\
& ... &      6.5  & - &   0.06$\pm$0.02  &  ...  & T14 \\
(136472) Makemake$^{**}$ & Scattered &  11.24$\pm$0.01 &  20.54/\textbf{22.48} &  0.03$\pm$0.01 &  -0.3 &  O07\\
 & ...  &  7.7710$\pm$0.0030 &  &  0.0286$\pm$0.0016 &  ... &  H09\\
 & ...  &  7.65 & - &  0.014$\pm$0.002 &  ... &  T10\\
 & ...  &  7.65 & - &  0.022$\pm$0.002 &  ... &  T13\\
1996~RQ$_{20}$$^{*}$ & Classical &  - &  - &  - &  6.9 &  -  \\  
(20161) 1996~TR$_{66}$$^{*}$ & Resonant &  - &  - &  - &  7.5 &  -  \\ 
1997~RX$_{9}$ & Classical &  - &  - &  - &  8.3 &  -  \\ 
1998~HL$_{151}$& Classical  &  - &  - &  - &  8.1 &  -  \\ 
(181855) 1998~WT$_{31}$ & Classical &  - &  - &  - &  7.2&  -  \\ 
1999~CD$_{158}$$^{*}$ & Resonant &  - &  - &  0.6 &  5.1 &  S10  \\ 
           & ...       &  -   &  6.88 &  0.49$\pm$0.03 & ...  &  TW  \\ 
(40314) 1999~KR$_{16}$ & Detached &  (5.840 or 5.929)$\pm$0.001&   (11.680 or 11.858)$\pm$0.002 &  0.18$\pm$0.04 &   5.8  &  SJ02  \\
        & ...            &  5.8   &   - &  0.12$\pm$0.06 & ... & T12  \\
1999~OH$_{4}$ & Classical &  - &  - &  - &  8.3 &  -  \\ 
1999~OK$_{4}$  & Classical&  - &  - &  - &  7.6 &  -  \\ 
(86177) 1999~RY$_{215}$$^{*}$& Classical&  - &   -  &  - &   7.1  &   -  \\
2000~CG$_{105}$$^{*}$& Classical&  - &   -  &  0.45 & 6.6  &   S10  \\
            & ...      & $>$2  & $>$4 &  $>$0.2  & ... &  TW \\ 
(130391) 2000~JG$_{81}$& Resonant &  - &   -  &  - & 8.0  &   -  \\
2001~FU$_{172}$ & Resonant &  - &   -  &  - & 8.3  &   -  \\
2001~QC$_{298}$ & Classical&  - &  $\sim$12 &  0.4 &  6.1 &  S10\\
& ...  & 3.89$\pm$0.24 & 7.78$\pm$0.48 & 0.30$\pm$0.04 & ...  & K06\\
(55565) 2002~AW$_{197}$ & Classical&  8.86$\pm$0.01 &   -  &  0.08$\pm$0.07 &   3.5  &   O06  \\
               & ...                        &  8.78 &   -  &  0.02$\pm$0.02 &   ...  &   T10  \\
2002~GH$_{32}$& Classical&  - &   -  &  0.75 &   5.5  &  S10  \\
               & ...   & -  & 3.98 &  0.36$\pm$0.02   & ... &  TW \\                 
2003~HA$_{57}$ & Resonant &  - &  - &  $\sim$0.5 & 8.1 &  R05  \\ 
               & ...   &  -  & 6.44 & 0.31$\pm$0.03   & ... &  TW \\ 
2003~HX$_{56}$ & Classical &  - &  - &  $\sim$0.4 & 7.1 & R05  \\ 
                          & ...      & -   & $>$10  &  $>$0.4  & ... &  TW \\ 
2003~QX$_{91}$  & Resonant  &  - &  - &  - &8.3&  -  \\ 
2003~TH$_{58}$ & Resonant&  - &   -  & $\sim$0.2 &  7.6  &   R05 \\
2004~PT$_{107}$ & Classical &  $\sim$20 &   -  &  0.05 &   6.0  &   S10  \\
2005~GE$_{187}$$^{*}$ & Resonant &  6.1 &  - &  0.5 &  7.1 &  S10\\
				 & ...  &  - &  11.99 &   0.29$\pm$0.02  &  ... &  TW\\
(202421) 2005~UQ$_{513}$& Classical &  7.03 or 10.01 &   -  &  0.06$\pm$0.02  &   3.4  &  T12  \\
2008~AP$_{129}$ & Classical &  9.04 &  - & 0.12$\pm$0.02 &  4.7 &   TW \\
2010~KZ$_{39}$& Scattered & - &- & $<$0.17 & 4.0 &  BS13 \\
2014~FT$_{71}$ & Classical & -  & -  & -  & 4.7  &  -  \\
\enddata
\\
Notes: \\
$^{*}$ No satellite/companion detected with \textit{Hubble Space Telescope}.\\
$^{**}$: different ligthcurve amplitudes according to filter used, as reported in \citet{Heinze2009}, and \citet{Thirouin2013}. \\
$^{a}$ Haumea's moons. \\
$^{b}$ contact binary \citep{Lacerda2014} \\

References list: \\
RT99: \citet{Romanishin1999};
H00: \citet{Hainaut2000};
SJ02: \citet{SheppardJewitt2002};
SJ03: \citet{SheppardJewitt2003};
R05: \citet{Rousselot2005};
O04: \citet{Ortiz2004};
B06: \citet{Belskaya2006}; 
K06: \citet{Kern2006_phd};
O06: \citet{Ortiz2006};
R06: \citet{Rabinowitz2006};
O07: \citet{Ortiz2007};
L08: \citet{Lacerda2008};
R08: \citet{Rabinowitz2008};
H09: \citet{Heinze2009};
P09: \citet{Perna2009};
S10: \citet{Snodgrass2010};
T10: \citet{Thirouin2010};
S12: \citet{Sonnett2012};
T12: \citet{Thirouin2012};
BS13: \citet{Benecchi2013};
T13: \citet{Thirouin2013};
L14: \citet{Lacerda2014};
T14: \citet{Thirouin2014};
TW: this work.
\end{deluxetable}

\begin{scriptsize}
\begin{table}
\caption{\label{Summary_photo} Summary of results from this work. In this table, we present the preferred rotational period (Rot. per. in hour), the preferred photometric period (Phot. per. in hour) and the peak-to-peak lightcurve amplitude ($\Delta$m in magnitude), the Julian Date ($\varphi_{0}$) for which the phase is zero in our lightcurves. The Julian Date is without light time correction. The preferred photometric period is the periodicity obtained thanks to the data reduction. Preferred rotational period is estimated based on our criteria to distinguish if a lightcurve is shape- or albedo-dominated and based on the asymmetry of the lightcurve (see Section~\ref{sec:singledouble} for more details). We also report if the lightcurve is asymmetric or not (Asym. LC column) and if the object is known to be a binary one (i.e. binary companion detected or not with the \textit{Hubble Space Telescope} (HST), last column). Some objects have not been observed with HST to detect binarity and are indicated with a question mark.  
}
 \begin{scriptsize}
\begin{tabular}{@{}lcccc|cc} 
\hline
 Object  & Phot. per. & Rot. per. & $\Delta$m &  $\varphi_{0}$ [JD] & Asym. &  Bin. \\
 
  & [h] &  [h]& [mag] & [2450000+] & LC? & ?  \\
\hline
\hline
1995~SM$_{55}$ & 4.04   & 8.08   & 0.04$\pm$0.02  &  2193.90249$^{a}$ &Yes & No \\
1999~OY$_{3}$ &   9.01  &  18.02   & 0.08$\pm$0.02  &  7167.88515  & Yes & No \\
2003~OP$_{32}$ & 4.85  & 4.85 & 0.14$\pm$0.02   &  3588.39312   &No &  No \\
2003~UZ$_{117}$ & 5.65 & 11.29  & 0.09$\pm$0.01 &  4438.54307$^{b}$ & Yes &  No    \\
2009~YE$_{7}$  & 5.65  & 5.65 &  0.06$\pm$0.02 &  5833.71848$^{c}$ & No &  ?  \\
2008~AP$_{129}$  & 9.04  & 9.04 & 0.12$\pm$0.02  &   5952.41458 & No &  ?  \\
1999~CD$_{158}$  & 3.44 & 6.88 &  0.49$\pm$0.03 &   7105.61716 &Yes & No  \\
2000~CG$_{105}$$^{*}$  & $>$2 & - & $>$0.2  &  7106.61325 & - & No \\
2002~GH$_{32}$  & 1.99 & 3.98 & 0.36$\pm$0.02  &   7106.59003 & Yes & ? \\
2003~HA$_{57}$  & 3.22 & 6.44 & 0.31$\pm$0.03   & 7105.86625  & Yes &  ?   \\
2003~HX$_{56}$$^{**}$  & $>$5 & $>$10 &  $>$0.4 &  7168.65047  & - &  ? \\
2005~GE$_{187}$  & 5.99 & 11.99 &  0.29$\pm$0.02 &   4622.47907$^{d}$& Yes &  No  \\
\hline\hline
\end{tabular}
\\
Notes:\\
$^{*}$: Amplitude variation based on $\sim$2~h of observations (see Section~\ref{sec:CG105})\\
$^{**}$: Constraints for lightcurve amplitude and rotational period. Based on the large amplitude, we favor the double-peaked option. \\
$^{a}$: Zero phase from \citet{SheppardJewitt2003}.   \\
$^{b}$: Zero phase from \citet{Perna2009}.   \\
$^{c}$: Zero phase from \citet{Benecchi2013}.   \\
$^{d}$: Zero phase from \citet{Snodgrass2010}.   \\
 \end{scriptsize}
\end{table}
\end{scriptsize}

\begin{table}
\caption{\label{Meanperiod} "STDEV" stands for Standard Deviation, "SE" for Standard Error and are associated to the average and median rotational periods. The last column report the mean rotational period obtained from the Maxwellian fits. All values are in hours. We consider several sub-samples to consider the confirmed family members, the candidates as well as the other TNOs. We also removed the binary population (labeled as binary pop) from the last sample because it has been shown that such a population is affected by tidal effects between the components of the system and therefore, their rotational period is not primordial (for more details see \citet{Thirouin2014}).     }
\begin{tabular}{@{}lcccc} 
\hline
 Sample  & Average & Median & SE/STDEV & Maxwellian       \\
 
 &     [h]& [h] & [h]  &[h]  \\
\hline
\hline
Confirmed$^{a}$ & 7.90 &  7.21  & 1.24/3.92 &   6.27$\pm$1.19   \\   
Candidates   & 8.09  & 6.88 & 1.43/4.53  &  6.44$\pm$1.16    \\   
Other TNOs,  &  13.85  & 8.55 & 8.24/26.06  &    8.98$\pm$0.59   \\
and candidates &    &  &  &    \\   
Other TNOs, no candidates   &  14.64  & 8.84 & 8.82/27.89  &    7.65$\pm$0.54   \\ 
Other TNOs, no candidates, &8.96&8.22&1.04/3.29&   8.98$\pm$0.77   \\  
no binary pop  &    &  &  &    \\
\hline\hline
\end{tabular}
\\
Note:\\
$^{a}$: Average is 6.89~h, median is 6.99~h and SE/STDEV are, respectively, 0.68/2.14
for the confirmed family members without the slow rotator 1999~OY$_{3}$.
\end{table}

\begin{table}
\caption{\label{tab:Density} Elongation and lower limit to the density for all objects studied are summarized in this table. We consider two cases: i) equatorial view (viewing angle of 90$^\circ$), and ii) viewing angle of 60$^\circ$. Lower limit to the density has been computed using \citet{Chandrasekhar1987}. As mentioned in the discussion, this model assumes that objects are in hydrostatic equilibrium and are triaxial (Jacobi) objects. Lower limits for MacLaurin spheroids are also reported here but, one has to keep in mind that these densities are based on assumptions that do not hold for this kind of objects.    }
 
\begin{tabular}{@{}lcc||cc} 
\hline
 Object  & b/a & b/a   & $\rho$ & $\rho$       \\
 
  &     eq.view & $\xi$=60$^{\circ}$ & eq.view   & $\xi$=60$^{\circ}$    \\
 &      &  & [g cm$^{-3}$]  & [g cm$^{-3}$]  \\
\hline
\hline
1995~SM$_{55}$ & 0.95 & 0.83  &   $>$0.60 & $>$0.60  \\
1999~OY$_{3}$ &  0.93 & 0.81 &   $>$0.12 &  $>$0.12  \\
2003~OP$_{32}$ &  0.88   & 0.76  &  $>$1.66  & $>$1.70  \\
2003~UZ$_{117}$ &  0.92   & 0.80  &  $>$0.31  &  $>$0.31 \\
2009~YE$_{7}$  &   0.95  & 0.82  &  $>$1.22  & $>$1.23  \\
2008~AP$_{129}$  &   0.90  &  0.78 &  $>$0.48  & $>$0.49  \\
1999~CD$_{158}$  &   0.63 & 0.54  &  $>$0.85  & $>$0.89  \\
2000~CG$_{105}$$^{*}$  & 0.83 & 0.72 &   	$>$0.61   &    	$>$0.63  \\
2002~GH$_{32}$  & 0.72 & 0.62 &  $>$2.56 &    $>$2.68 \\
2003~HA$_{57}$  &0.75  & 0.65 &  $>$0.87 &   $>$1.01  \\
2003~HX$_{56}$$^{**}$   & 0.69 & 0.60 &  $>$0.41&   $>$0.43  \\
2005~GE$_{187}$  & 0.63 & 0.55 &  $>$0.89 &   $>$0.94  \\
\hline\hline

\end{tabular}
 \newline
Notes:\\
$^{*}$: Axis ratio derived assuming a lightcurve with a 0.2~mag as amplitude. Assuming a rotational period of 8~h for this object (mean rotational period of the non-binary TNOs as reported in \citet{Thirouin2014}), we derived a lower limit to the density $\sim$0.6~g cm$^{-3}$. \\
$^{**}$: Axis ratio derived assuming a lightcurve with a 0.4~mag as amplitude. Assuming a rotational period of 10~h for this object, we derived a lower limit to the density $\sim$0.4~g cm$^{-3}$. \\

\end{table}

\begin{deluxetable}{lccccc}
\tablewidth{0pt}
\tablecaption{\label{tab:HaumeaMembers} We summarize the diameter, the mass, the albedo and the absolute magnitude for confirmed, and candidates. Absolute magnitudes (H) are from the Minor Planet Center (MPC) database. Diameters reported in this table are only approximations. More accurate diameters and albedos for some family members will be published in Vilenius et al. (In prep) using thermal modelling of \textit{Herschel Space Observatory} data.   }
\tablehead{
\colhead{Object}  & \colhead{H}  & \colhead{Albedo$^{a}$}   & \colhead{Diameter [km]} & \colhead{Mass$^{b}$ [$\times$10$^{18}$~[kg]]} & \colhead{Ref.}}
\startdata
\textit{Confirmed members:} &    &  &   &  &   \\
1995~SM$_{55}$ & 4.8 & 0.70 &	149 & 1.73 & TW\\
1996~TO$_{66}$ & 4.5 & 0.70 &	171 & 2.62 & TW\\
1999~OY$_{3}$ & 6.8 & 0.70 &	59 &  0.11 &TW\\
2002~TX$_{300}$ & 3.2 &  0.88$^{+0.15}_{-0.06}$  & 286$\pm$10  & 12.25&  E10\\
Haumea  & 0.1 & 0.70 - 0.75 &	1150  & 4006$\pm$40 & L10, R09\\
2003~OP$_{32}$ & 3.6 & 0.70 &	259 & 9.09 &TW\\
2003~SQ$_{317}$ & 6.3 & 0.70 & 75 & 0.22 &TW\\
2003~UZ$_{117}$ & 5.3 & 0.70 & 118 & 0.87 &TW\\
Hi'iaka  & 2.9 & $\sim$0.70 &	$\sim$320  & 17.9$\pm$1.1 &R09\\
Namaka  & 4.5 & $\sim$0.70 &	$\sim$160 & 1.79$\pm$1.48&R09 \\
2005~CB$_{79}$ & 4.7 & 0.70 &	156 & 1.99 &TW\\
2005~RR$_{43}$ & 4.0 & 0.70 &	215 &  5.23 &TW\\
2009~YE$_{7}$ & 4.4 & 0.70 &	179 & 3.01 &TW\\ 
\hline
\textit{Candidates:}  &    &  &   &  &   \\
Salacia & 4.4 & 0.0439$\pm$0.0044 & 901$\pm$45 & 311.96$\pm$46.71 & V12\\ 
Makemake & -0.3 & 0.77$\pm$0.03 & 1430$\pm$9  &2761.05$\pm$161.26  &  O12 \\ 
1996~RQ$_{20}$ & 6.9 & 0.08/0.30 &  168/87 & 2.46/0.34 & TW\\ 
1996~TR$_{66}$  & 7.5& 0.08/0.30 &127/66& 1.08/0.15 &TW\\  
1997~RX$_{9}$   & 8.3&  0.08/0.30 & 88/45& 0.36/0.05 & TW\\ 
1998~HL$_{151}$     & 8.1& 0.08/0.30 & 96/50 & 0.47/0.06 &TW\\  
1998~WT$_{31}$  & 7.2& 0.08/0.30 & 146/75 & 1.63/0.22 & TW\\ 
1999~CD$_{158}$   & 5.1& 0.08/0.30 & 384/198 & 29.62/4.08 &TW\\  
1999~KR$_{16}$    & 5.8 & 0.204$_{-0.05}^{+0.07}$ & 254$\pm$37 & 4.79$\pm$1.95 & SS12 \\ 
1999~OH$_{4}$      &8.3 &0.08/0.30 & 88/45 &0.36/0.05 & TW\\ 
1999~OK$_{4}$     & 7.6&  0.08/0.30 &121/63 &0.94/0.13 &TW\\  
1999~RY$_{215}$    &7.1 & 0.0388$^{+0.0122}_{-0.0065}$ & 263$^{+29}_{-37}$& 8.75$\pm$2.89  & V14\\ 
2000~CG$_{105}$   &6.6 &  0.08/0.30 &192/99 &  3.73/0.51 &TW\\  
2000~JG$_{81}$   & 8.0&  0.08/0.30 & 101/52 & 0.54/0.07 & TW\\ 
2001~FU$_{172}$  & 8.3&  0.08/0.30 & 88/45& 0.34/0.05 &TW\\  
2001~QC$_{298}$   &6.1 & 0.061$^{+0.027}_{-0.017}$ & 303$^{+27}_{-30}$&  19.72$\pm$9.42 & V14\\ 
2002~AW$_{197}$  & 3.5 & 0.112$^{+0.012}_{-0.011}$ & 768$^{+39}_{-38}$& 263.87$\pm$40.29 & V14\\ 
2002~GH$_{32}$$^{c}$   & 5.5& $>$0.13 & $<$180 & $>$13.14  & V14\\ 
2003~HA$_{57}$    & 8.1& 0.08/0.30 &96/50 & 0.47/0.06 & TW\\ 
2003~HX$_{56}$    & 7.1& 0.08/0.30 &153/79& 1.87/0.26 & TW\\
2003~QX$_{91}$      & 8.3&  0.08/0.30 & 88/45 &0.34/0.05 & TW\\ 
2003~TH$_{58}$   & 7.6&  0.08/0.30 &121/63 &0.94/0.13 & TW\\ 
2004~PT$_{107}$   & 6.0& 0.0325$^{+0.0111}_{-0.0066}$ & 400$^{+45}_{-51}$& 53.99$\pm$20.08  & V14\\ 
2005~GE$_{187}$   & 7.1&  0.08/0.30 & 153/79 & 1.87/0.26 & TW\\
2005~UQ$_{513}$   & 3.4& 0.202$^{+0.084}_{-0.049}$ & 498$^{+63}_{-75}$& 130.30$\pm$57.01 & V14\\ 
2008~AP$_{129}$ & 4.7 & 0.08/0.30 & 462/238 & 51.48/7.09 & TW\\ 
2010~KZ$_{39}$   &4.0 &  0.08/0.30 & 637/329 & 135.41/18.65 & TW\\
2014~FT$_{71}$   & 4.7 & 0.08/0.30  & 462/238 & 51.48/7.09 & TW \\
\hline
\enddata
 \newline
Notes:\\
$^{a}$: Assuming an albedo of 0.70 for confirmed members of the family, except for Haumea and 2002~TX$_{300}$ whose albedos are known. Assuming albedos of 0.08/0.30 for the candidates when the albedo is unknown.   \\
$^{b}$: Masses (except Haumea, Hi'iaka and Namaka masses) computed assuming a density of 1~g~cm$^{-3}$.  \\
$^{c}$: Assuming an albedo of 0.08/0.30, we derived a diameter of 319/165~km and a mass of 17.05/2.35$\times$10$^{18}$~kg. \\
References:\\
R09: \citet{Ragozzine2009}, E10: \citet{Elliot2010}, L10: \citet{Lellouch2010}, O12: \citet{Ortiz2012Makemake}, SS12: \citet{Santos-Sanz2012}, V12: \citet{Vilenius2012}, V14: \citet{Vilenius2014}, TW: this work. 
\end{deluxetable}

\begin{deluxetable}{lcccc}
\tablewidth{0pt}
\tablecaption{\label{Tab:Correlations} Some correlations/anti-correlations found using the lightcurve parameters and orbital/physical variables. We looked into thirteen samples of data. \textit{family} stands for all confirmed members of the family, \textit{candidates} are reported in Table~\ref{Tab:allphoto}, \textit{Other TNOs} refers to all TNOs without relation with the family, and \textit{no binary pop} excludes the binary population. We indicate the Spearman rank correlation ($\rho$), the Significance Level (SL in percent), and the number of objects in each sample (Nb). Only positive correlations/anti-correlations with a Spearman rank and Significance Level in agreement with our criterion are reported.}
\tablehead{
\colhead{Correlated values}  & \colhead{Sample}  & \colhead{$\rho$}   & \colhead{SL [\%]} & \colhead{Nb.} }
\startdata
Rotational period versus absolute magnitude  &  Family  & 0.638  & 98.43 &  11  \\
Rotational period versus inclination  &  Family  & -0.565 & 96.33 &  11  \\
Rotational period versus ascending node  & Family  & 0.426 & 95.27  &  11  \\
                                        &  Family, and candidates    & 0.358  & 99.28 &  23  \\
Lightcurve amplitude versus absolute magnitude  & Family, and candidates  & 0.473  & 99.32 &  23 \\
\enddata
\end{deluxetable}
\clearpage

\clearpage

\end{document}